\documentclass{aa}  

\usepackage{hyperref}

\usepackage{graphicx}
\usepackage{subfigure}
\usepackage{booktabs}

\usepackage{threeparttable}
\usepackage{txfonts}

\begin{document} 

   \title{What drives the wheels of evolution in NGC 1512?}

   \subtitle{A UVIT study}

   \titlerunning{A higher resolution UV view of NGC 1512}
\authorrunning{Robin et al.}

   \author{T. Robin\inst{1}\thanks{e-mail: robin.thomas@christuniversity.in}, 
           Sreeja S Kartha \inst{1},
          Ujjwal Krishnan$^{1}$,
         Kanak Saha$^{2}$,
          Viral Parekh$^{3}$,
          Koshy George$^{4}$,
        Blesson Mathew$^{1}$\\}
    \authorrunning{Robin et al. }
   \institute{Department of Physics and Electronics, CHRIST (Deemed to be University), Hosur Main Road, Bangalore, India
   \and Inter-University Centre for Astronomy and Astrophysics, Ganeshkhind, Post Bag 4, Pune 411007, India
   \and National Radio Astronomy Observatory (NRAO), 1003 Lopezville Rd, Socorro, NM 87801, USA
   \and Faculty of Physics, Ludwig-Maximilians-Universität, Scheinerstr 1, Munich, 81679, Germany}

   \date{Received xxx; accepted xxx}

 
  \abstract
    { Environmental and secular processes play a pivotal role in the evolution of galaxies. These can be due to external processes such as interactions or internal processes due to the action of bar, bulge and spiral structures. {Ongoing star formation} in spiral galaxies can be affected by these processes. By studying the star formation progression in the galaxy, we can gain insights into the role of {different processes that regulate the overall evolution} of a galaxy.}
   { 
   The ongoing interaction between barred-spiral galaxy NGC 1512 and its satellite NGC 1510 {offers an opportunity to investigate} how galactic interactions and the presence of a galactic bar influence the evolution of NGC 1512. {We aim to understand the recent star formation activity in the galaxy pair and thus gain insight into the evolution of NGC 1512}. }
   { {The UltraViolet Imaging Telescope (UVIT)} onboard \textit{AstroSat} enables us to characterize the star-forming regions in the galaxy with a superior spatial resolution  of $\sim$ 85 pc {in the galaxy rest frame}. We identified and characterized 175 star-forming regions in the UVIT FUV image of NGC 1512 and correlated with the neutral hydrogen (H\textsc{i}) distribution. {Extinction correction was applied to the estimated} photometric magnitude. We traced the star-forming spiral arms of the galaxy and studied the star formation properties across the galaxy in detail.}
   { We detected localized regions of star formation enhancement and distortions in the galactic disk. We found this to be consistent with the distribution of H\textsc{i} in the galaxy. {This is evidence of past} and ongoing interactions affecting the star formation properties of the galaxy. We studied the properties of the inner ring. We find that the regions of the inner ring show maximum star formation rate density ($\mathrm{log(SFRD_{mean} [M_{\odot}yr^{-1}kpc^{-2}]) \sim -1.7}$) near the major axis of the bar, {hinting at a possible} crowding effect in these regions. The region of the bar in the galaxy is also depleted of UV emission. This absence suggests that the galactic bar may have played an active role in the redistribution of gas and quenching of star formation {inside identified bar region}. Hence, we suggest that both the secular and environmental factors might be {influencing the evolution of NGC 1512.}}
   {}


   \maketitle
%

\section{Introduction}

Galaxies are fundamental building blocks of the Universe. Understanding the evolutionary processes of galaxies requires a comprehensive investigation of the interplay between environmental effects, such as interactions, and internal structures like bars, oval disks, and spiral arms. In the context of the Local Universe, there is a transition occurring in the primary drivers of galactic evolution. In the early Universe, the dominant forces were characterized by violent and rapid processes, such as mergers. However, there is now a discernible shift towards a more gradual restructuring of mass and energy, attributable to the influence of secular evolution \citep{Kormendy2004ARA&A..42..603K}. Decoding the aspects of a galaxy, such as its gas content, star formation activity, and structural properties, provides valuable insights into its evolution. This understanding can be achieved by recognizing the significant impacts exerted by both secular and environmental factors on these crucial galaxy characteristics \citep{toomre1972galactic, Skibba2009MNRAS.399..966S}.

One of the observable parameters that can be used to understand the factors influencing the evolution of a galaxy is the star formation activity. By measuring the star formation rate (SFR) and analyzing the spatial distribution of young stars, crucial insights into the galaxy's evolution can be obtained. The intense ultraviolet continuum radiation emitted by young massive stars of spectral types O, B, and A serves as a direct indicator of ongoing and recent star formation \citep{Kennicutt1998}. The understanding of star formation goes beyond quantifying the star formation rate (SFR) and requires an exploration of the underlying mechanisms governing this process. At its core, star formation relies on the presence of cold, neutral gas as the fundamental ingredient. The significance of neutral hydrogen (H\textsc{i}) gas in supporting and sustaining star formation is well-established \citep{Doyle2006MNRAS.372..977D,Leroy2008AJ....136.2782L,Kennicutt2012ARA&A..50..531K,Zhou2018PASP..130i4101Z, Parkash2018ApJ...864...40P}. By examining the UV properties of galaxies and analyzing the spatial distribution of H\textsc{i} gas, valuable insights can be gleaned regarding the mechanisms responsible for recent star formation and, consequently, the evolution of the galaxy. This approach allows for a comprehensive understanding of the interplay between star formation processes and the presence of neutral hydrogen gas, shedding light on the underlying dynamics shaping galactic evolution.

\begin{figure}
\begin{center}
{\includegraphics[width = 1\columnwidth]{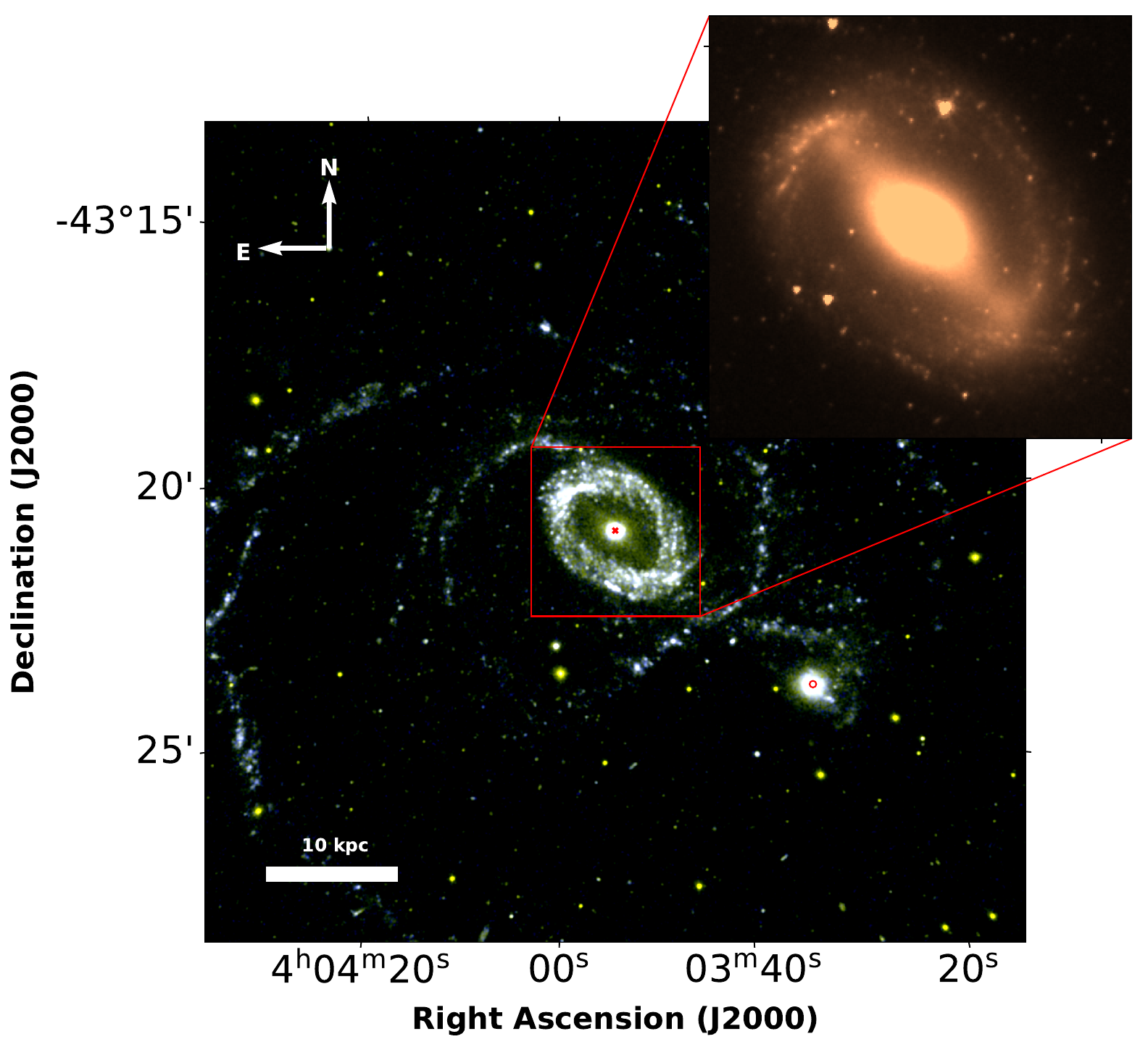}}
\caption{UVIT colour composite image of the galaxy pair NGC 1512/1510. {The galaxy in the centre is NGC 1512, as indicated by the red cross. The satellite galaxy, NGC 1510, is identified by the red circle and is at a distance of $\mathrm{\sim 5 \arcsec (18.3kpc)}$ from NGC 1512}. The emission in F154W and N242W are represented by blue and yellow colours, respectively. The yellow points in the image are foreground stars, confirmed using {the} \textit{Gaia} DR3 catalogue \citep{Gaia2022arXiv220800211G}. The inset shows IRAC 3.6$\mu$m image of the galaxy, with the galactic bar visible. }

\label{fig:composite}
\end{center}
\end{figure}

{The galaxy pair} NGC 1512/1510 is located on the edge of the Local Volume at a distance of 12.60 Mpc \citep{TullyDISTANCE2016AJ....152...50T}. The massive barred-spiral galaxy NGC 1512 and its satellite, NGC 1510, a Blue Compact Dwarf (BCD), separated by a distance of $\sim$ 14 kpc, are undergoing interaction that started about 400 Myr ago \citep{Koribalski_1512_2009}. NGC 1512 hosts a {strong bar \citep[bar strength, $\mathrm{Q_{b}} \sim 0.27$, ][]{Buta2005AJ....130..506B,Buta2006AJ....132.1859B}}, which {might have} formed as a result of interactions {\citep{Koribalski_1512_2009}}. It hosts a circumnuclear ring of diameter $\sim$ 16\arcsec $\times$ 12\arcsec and is subjected to intense starbursts. The galaxy also hosts an {oval} inner ring of diameter 3\arcmin $\times$ 2\arcmin, which is elongated along the major axis of the bar (with the relative angle between the bar and inner major axis $\gtrsim$ 10$^\circ$) \citep{Maoz2001AJ....121.3048M,Meurer2006ApJS..165..307M}. The inner ring is notable for its sharp definition due to an enhanced degree of star formation \citep{Hawarden1979A&A....76..230H}. The galaxy has spiral arms wound around it, with one of the arms disrupted, possibly due to its interaction with the NGC 1510. Studies have also discussed the possibility of arms being formed due to tidal interactions \citep{Kinman_arms1978AJ.....83..764K, Ducci2014A&A...566A.115D}. The properties of the galaxy pair are listed in Table \ref{tab:parameter}. We identify NGC 1512 as the main galaxy and NGC 1510 as the satellite galaxy, based on their mass ratio of 50:1 \citep{Chakrabarti2011ApJ...743...35C}. The galaxy NGC 1512 is, thus, an excellent target for our investigation, as it offers a compelling opportunity to study the {combined effects of environmental and secular components}. 
\begin{table}
  \centering
  \begin{threeparttable}
  \caption{Basic parameters of NGC 1512/1510}
   \begin{tabular}{lccr} 
  \toprule
      \addlinespace[1ex]
  Parameter & NGC 1512 & NGC 1510 \\

      \addlinespace[1ex]
    \bottomrule
        \addlinespace[1ex]
  Type   & SB(r)a\tnote{1} &  SA0 pec - BCD\tnote{1} \\
  Central RA [hh mm ss] & 04 03 54.28\tnote{1}  &  04 03 32.6\tnote{1}\\
  Central Dec [dd mm ss]  & -43 20 55.9\tnote{1}& -43 24 00.5\tnote{1}\\
Redshift (z)  & 0.00299\tnote{2}& 0.00304\tnote{3}\\
  Inclination (i) & 42.5$^{\circ}$\tnote{4}&  47$^{\circ}$\tnote{1} \\
  Position angle (PA) & 40$^{\circ}$\tnote{4} &  90$^{\circ} $\tnote{1}\\
  \hline
  \end{tabular}
\begin{tablenotes}
    \item $^{1}$\citet{deVaucouleurs1991},
    $^{2}$\citet{Koribalski2004AJ....128...16K},$^{3}$\citet{Mathewson1996ApJS..107...97M},
    $^{4}$\citet{Maeda2023ApJ...943....7M}
\end{tablenotes}
\label{tab:parameter}
 \end{threeparttable}
\end{table}

NGC 1512 possesses a unique combination of location, morphological features, and environmental characteristics that make it an ideal candidate for unraveling the role played by various evolutionary mechanisms in the local Universe. The presence of a bar in NGC 1512 introduces unique dynamics that profoundly impact the galaxy's structure and evolution, while the close proximity of the galaxy pair offers a unique opportunity to investigate the local impact of the interaction. Limited attention has been given to understanding how the presence of a galactic bar and the interactions with neighboring galaxies contribute to the evolution of NGC 1512 \citep{Hawarden1979A&A....76..230H, Kinman_arms1978AJ.....83..764K, Li2008,Koribalski_1512_2009,Ma2017ApJS..230...14M,Smirnova2020AstBu..75..234S}. {GALEX \citep{martingalex2005} has observed the galaxy pair NGC 1512/1510 with a resolution of 5\arcsec. However, the UltraViolet Imaging Telescope (UVIT) onboard \textit{AstroSat} \citep{Kumar2012SPIE.8443E..1NK} provides higher resolution data (resolution of 1.4\arcsec).} Thus, we aim to leverage the high-resolution UV data obtained from the UVIT, along with H\textsc{i} data from {MeerKAT \citep{Jonas2016mks..confE...1J, Jonas:2018Jr}}, {to gain insights into the evolution of NGC 1512}.

{This study investigates} the effects of interaction events and the galactic bar on the evolution of NGC 1512, with the aim of understanding how each of these factors influences star formation activity and galactic morphology. The paper is arranged as follows. The data and analysis are discussed in Section \ref{Data}. The results and discussion are presented in Section \ref{sec:results}. The summary is given in Section \ref{summary}. We have adopted a flat universe cosmology throughout this paper with $\mathrm{H_0}$ = 71 kms$^{-1}$Mp$c^{-1} $ and $\mathrm{\Omega_M }= 0.27$ \citep{Komatsu2011ApJS..192...18K}. In the galaxy rest frame, 1\arcsec corresponds to a distance of 60.9 pc.

\section{Data and Analysis}\label{Data}

\subsection{UVIT Data}
To understand the driving mechanisms behind the evolution of the interacting galaxy pair NGC 1512/1510, we use data from {UVIT, which is available at the \textit{AstroSat} ISSDC archive\footnote{https://astrobrowse.issdc.gov.in/astro\_archive/archive/Home.jsp}}. UVIT has three bands - Far UltraViolet (FUV; 130-180 nm), Near UltraViolet (NUV; 200-300 nm) and VISible (VIS; 320-550nm), with the capability to observe simultaneously in all three bands. The VIS channel is used to track the drift of the satellite during observations. The instrument has a 28$\arcmin$ field of view, spatial resolution of  $\sim1.4\arcsec$ and $ 1.2\arcsec$ for FUV and NUV filters, respectively and a plate scale of $\sim$ 0.416\arcsec pixel$^{-1}$. {The UVIT, thus, provides a better spatial resolution in UV when compared to its predecessor, GALEX.}

UVIT observed the galaxy pair of NGC 1512/1510 in FUV and NUV channels (PI: Kanak Saha, Obs. ID: $\mathrm{G07\_068}$). NGC 1512 has been observed in broadband filters of F154W  (FUV, hereafter) and N242W (NUV, hereafter). The observation details are {presented} in Table \ref{tab:obser}. We used the software package CCDLab \citep{Postma_Leahy_2017} to reduce the Level 1 data. Drift correction has been accounted for using the VIS images. Each image is corrected for fixed pattern noise, distortion and drift, and flat fielded using CCDLab \citep{Girish2017ExA....43...59G,Postma_Leahy_2017}. Final deep images are produced by combining the corrected images. The astrometric solutions are also made using the same software. Figure \ref{fig:composite} represents the UVIT colour composite image of NGC 1512/1510 pair generated using the emission in FUV and NUV. The zero point magnitude values for the FUV and NUV filters are 17.77 and 19.76 mag respectively \citep{Tandon2017,Tandon2020AJ....159..158T}. {The inset shows the IRAC} 3.6 $\mu$m image of the galaxy pair, where the galactic bar in NGC 1512 is visible.

\begin{table}
    \centering
    \caption{Log of observations}
    \begin{tabular}{lllll}
        \toprule
      \addlinespace[1ex]
        Facility &Filter& Peak $\lambda$ & Date of Obs. & Exp. Time \\ 
        
      \addlinespace[1ex]
      \bottomrule
          \addlinespace[1ex]
      UVIT & FUV & 1541 \AA & 30 Aug 2017 & 3276.2s \\
          & NUV & 2418 \AA &30 Aug 2017 & 3659.2s \\
        \hline
            \addlinespace[1ex]
        MeerKAT& &21 cm & 29 May 2019 & 21600s\\
        \hline
    \end{tabular}
    \label{tab:obser}
\end{table}

\subsection{MeerKAT Data}
The neutral hydrogen gas (H\textsc{i}) is a good tracer of gas distribution and kinematics in a galaxy. Previous H\textsc{i} observations of nearby galaxies have allowed the investigations of the possible {triggering mechanisms of starbursts} \citep{Lelli2014MNRAS.445.1694L,Zhang2020ApJ...900..152Z}. We obtained MeerKAT  {\citep{Jonas2016mks..confE...1J, Jonas:2018Jr}} L-band raw data from the archive \footnote{https://apps.sarao.ac.za/katpaws/archive-search} (Obs. ID: 20190515-0022) for NGC 1512/1510. The log of observations for the galaxy pair is tabulated in Table \ref{tab:obser}. At a frequency of 1.4 GHz, the telescope has a large field of view of 0.85 deg$^2$, covering the galaxy pair. The source J0408-6545 was employed as the flux density or primary calibrator, while the source J0440-4333 was the gain calibrator. We carried out the data reduction using CARACal \citep{Jozsa2020ASPC..527..635J}, which included both continuum and line imaging. We followed the standard data reduction procedure by removing bad channels, automatic removal of radio frequency interference (RFI; \citealp{offringa-2012-morph-rfi-algorithm}), and carrying out flux, bandpass and gain calibration.  {We used the SoFiA-2} \citep{Serra2015MNRAS.448.1922S} tool for detecting  H\textsc{i} sources and generating the clean mask used in WSClean \citep{Offringa2014MNRAS.444..606O} cube imaging. Further details on the data calibration, imaging, and source finding are summarised in \citet{Healy2021A&A...654A.173H}. In this paper, we only utilise MeerKAT data to correlate the distribution of H\textsc{i} with UV distribution for the galaxy pair. A detailed analysis to understand the H\textsc{i} dynamics in the galaxy pair is in preparation.

\subsection{Extinction}\label{sec:Extinction}

One of the major drawbacks while studying star formation in the UV continuum is its sensitivity to extinction \citep{Kennicutt1998}. {The UV flux determined should therefore be} corrected for extinction. We take the rest-frame extinction into account for our studies. {The reddening coefficients}, c($\mathrm{H\beta)_{internal}}$, is defined as the ratio of the observed flux density of $\mathrm{H\beta}$ to the unreddened flux density of the same line. This gives us an estimation of reddening caused by the interstellar dust. We crossmatched the star-forming regions identified in \citet{Lopez10.1093/mnras/stv703} with the identified star-forming regions in this study for reddening coefficient values. In case of a direct match, we adopted the reddening coefficient values for the identified region. For the regions identified newly in our study, we assumed an average value of {the reddening coefficient of the five nearest regions} identified in the previous study. We estimated the extinction in FUV by utilizing the extinction value of $\mathrm{A_v = 0.029}$ reported by \citet{Schlafly2011} for NGC 1512 and NGC 1510. The total reddening is calculated by taking the sum of rest-frame and Milky Way reddening values, as discussed in \citet{Karthick2014MNRAS.439..157K}:

\begin{equation}
     \mathrm{E(B - V)_{Total} =  E(B - V)_{MW} + 0.692 \times  c(H\beta)_{internal}}
\end{equation}

We used the \citet{Cardelli1989ApJ...345..245C} extinction law with $\mathrm{R_v = (A_V)/ E(B-V)} = 3.1$ and utilised the extinction module in Astropy \citep{Astropy2013A&A...558A..33A} to estimate the value of extinction coefficients ($\mathrm{R_\lambda}$) for filters of FUV and NUV bands for individual regions. After estimating the total extinction value, we corrected the FUV and NUV magnitudes, which will be used in further analysis.

\subsection{{Identification of star-forming regions in NGC 1512}}

Star-forming regions provide important information about the evolutionary sequence of a galaxy. {We made use of the ProFound package to identify the brightest regions from the UVIT FUV band images}. ProFound is an astronomical data processing tool available in the R programming language \citep{Robotham_Davies_Driver_Koushan_Taranu_Casura_Liske_2018}. Using watershed deblending, ProFound locates the image's peak flux areas and identifies the source segments. The total photometry is then estimated using iterative expansion (dilation) of the observed segments. {Considering the spatial resolution of UVIT FUV filter ($\sim$ 1.4\arcsec, \citealp{Borgohain2022Natur.607..459B}), we defined a criterion that the identified regions should cover at least 6 pixels (the minimum number of pixels to cover the a circle with diameter as the spatial resolution of the FUV filter).} A \textit{skycut} of 3 was applied to identify star-forming regions. Details on source identification and background estimation are found in \citet{Robotham_Davies_Driver_Koushan_Taranu_Casura_Liske_2018} and \citet{Ujjwal2022MNRAS.tmp.2176U}. We identified 241 bright regions in the UVIT FUV image associated with the main galaxy NGC 1512.

\begin{figure*}[htbp]
  \centering
  \subfigure{\includegraphics[width=0.9\columnwidth]{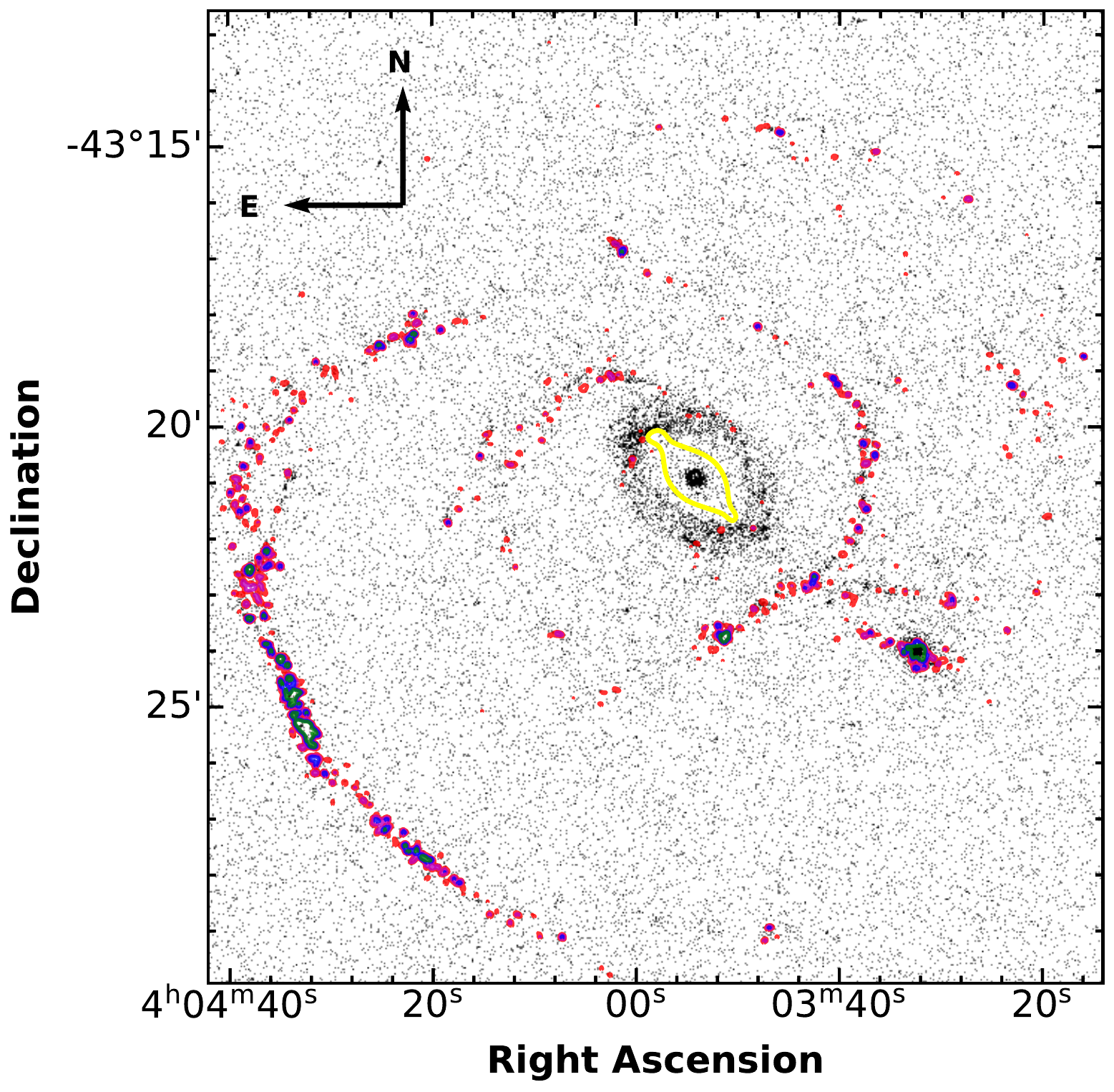}}
  \hfill
  \subfigure{\includegraphics[width=0.9\columnwidth]{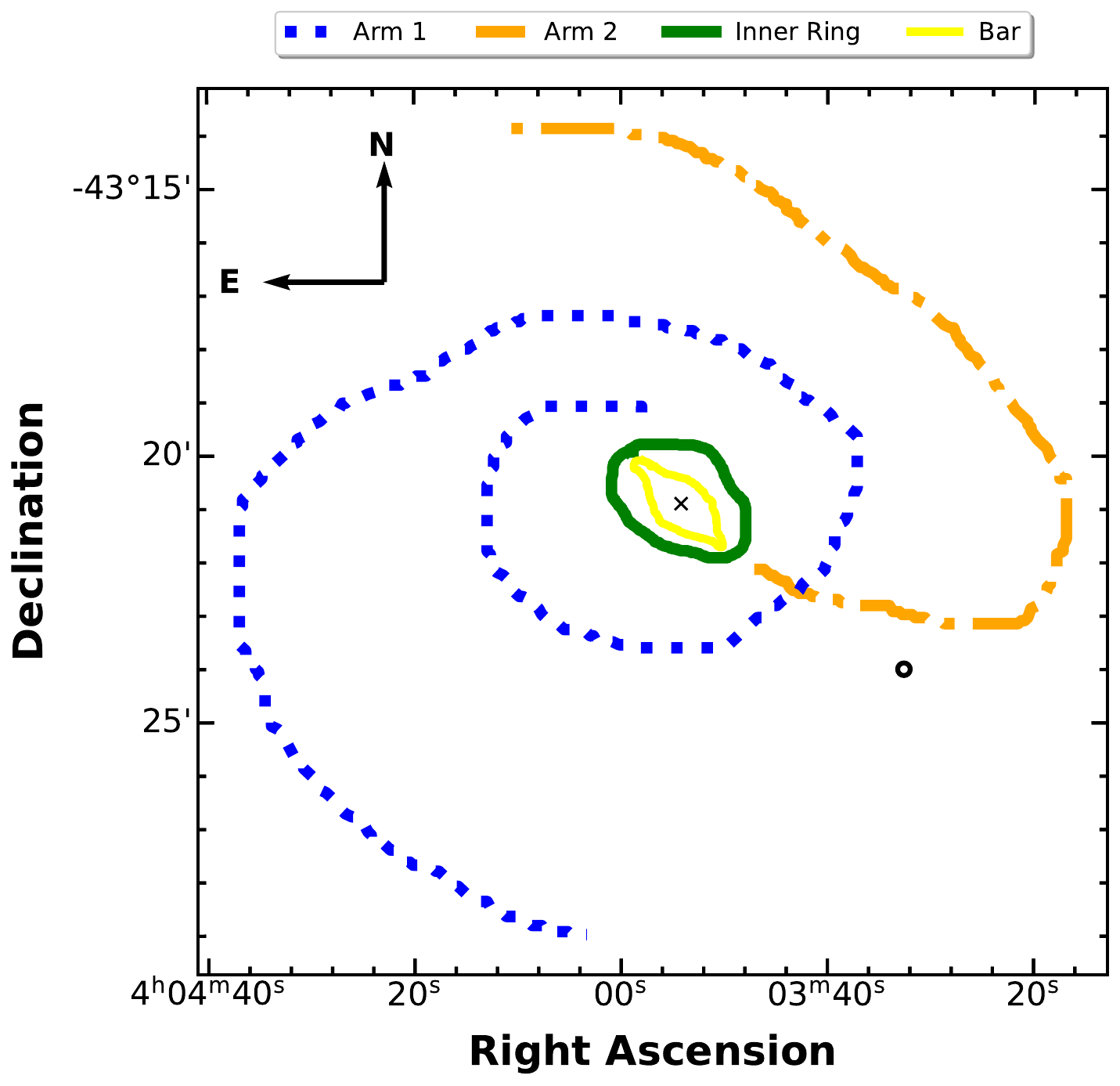}}
  \caption{(Left) FUV image of NGC 1512 overplotted with MeerKAT H\textsc{i} column density contours. The contours of red, magenta, blue and green colours represent column density values $\mathrm{(0.3, 0.35, 0.4\, and\, 0.45)\times  10^{21} cm^{-2}}$ respectively. We {extracted the contour over the galactic bar from the IRAC 3.6$\mu$m image to identify its extent and overplotted it in yellow on the FUV image.}
  (Right) A {schematic diagram} of the morphologies identified in NGC 1512. The blue dotted line represents Arm 1, the orange dash-dotted line represents Arm 2, and the green continuous line represents the inner ring. The galaxy centres for NGC 1512 and NGC 1510 are represented with a cross mark and open circle, respectively.}
  \label{fig:morph}
\end{figure*}

\subsection{{Identification of spiral arms in NGC 1512}}

We observe the presence of a bright nuclear region and an inner ring in NGC 1512. We try to understand the morphology of the spiral arms in the galaxy. Previous studies on NGC 1512 noted a single spiral arm beginning at the North-East edge of the inner ring and wrapping continuously around NGC 1512 for an azimuthal angle of nearly 540$^{\circ}$ \citep{Thilker2007ApJS..173..538T,Bresolin2012}.

To understand and segregate the identified regions into different arms, we estimated the deprojected distance of the regions along the line of sight {by utilising the equations (1-11) provided in Section 2} of \citet{Marel2001AJ....122.1807V}. We observed that the regions arching from the NE of the inner ring of the main galaxy towards the satellite galaxy exhibit a lesser distance along the line of sight compared to the regions in the SW of the inner ring and in the immediate vicinity of the satellite galaxy. {The estimated distance difference between the regions originating from NE of the inner ring and those originating from SW of the inner ring was of the order of $\sim$ 20 kpc}. The MeerKAT H\textsc{i} image of the galaxy also showed a similar trend. The regions originating from the NE of the inner ring have a continuous gas distribution and {wind around the galaxy}, while the regions in the SW of the inner ring and near the satellite galaxy exhibit a distorted H\textsc{i} distribution that extends outwards, as is visible in Figure \ref{fig:morph} (left panel).

Keeping this distribution in mind, we identify the spiral arms of the galaxy as follows. Arm 1 originates from NE of the inner ring and traces the galaxy for 540$^{\circ}$ azimuthally and Arm 2 originates from the SW region of the inner ring (Figure \ref{fig:morph}, right panel). Arm 2 undergoes disruption due to the interaction between the galaxy pair. It also shows a tidal bridge-like feature that extends from the main galaxy towards the satellite galaxy, following which it extends outwards. The regions identified as part of Arm 1 have been represented with a dotted line, whereas Arm 2 has been represented with a dash-dotted line. We also detected the galactic bar in NGC 1512 using IRAC 3.6$\mu$m image (PI: Robert Kennicutt). The identified bar region is shown in yellow colour in the left panel of Figure \ref{fig:morph}. {The half-length of the bar is measured to be $\sim73.5\arcsec$ (4.4 $\pm$ 0.33 kpc) by using the analysis from \citet{Peters2018MNRAS.476.2938P}}. Compared to the previous studies, UVIT allows us to study the star-forming regions individually and, thus, to study the possible effect of interaction in detail. 

\section{Results and Discussion}\label{sec:results}

\subsection{Estimation of SFR}\label{identity}

{The Profound module provides the number of pixels containing $100\%$ of the flux for all the identified 241 regions.} Using this, we calculated the area of each region. {The counts for each region were} also obtained from the same. The counts in the selected regions were integration time-weighted and converted to magnitude units using the zero point conversion factor, as discussed in \citet{Tandon2020AJ....159..158T}. {We corrected for extinction in the obtained magnitudes, in accordance with Section \ref{sec:Extinction}. We removed star-forming regions with {dereddened} FUV magnitude, $\mathrm{m_{FUV}}$, fainter than 21 mag to exclude regions with photometric errors larger than 0.1 mag \citep{Devaraj2023ApJ...946...65D}. We obtained a final sample of 175 star-forming regions in {NGC 1512}. Note that with the exception of analysis in Section \ref{sec:1510sfr}, regions of NGC 1510 have not been considered.} The corresponding star formation rates are then calculated using equation 4 in \citet{Kaisina2013} and is as follows.
\begin{equation}
    \mathrm{log(SFR[M_{\odot} yr^{-1}]) = 2.78 - 0.4 \times m^{c}_{FUV} + 2 \times log D}
\end{equation}

where SFR is the star formation rate, $\mathrm{m^{c}_{FUV}}$ denotes the total extinction corrected FUV magnitude, and D is the distance to the galaxy in Mpc. {The error in counts was estimated using Poisson's distribution. The errors associated with $\mathrm{m^{c}_{FUV}}$ and SFR were obtained from error propagation \citep{Bevington1992drea.book.....B}.}

To understand the effects of different galactic properties on determining the star formation rate and the {propagation of star formation throughout the galaxy}, we have classified the regions according to their position in the galaxy. We made use of the values of distance, the coordinates for the centre of the main galaxy, inclination, and position angle for the galaxy from Table \ref{tab:parameter} and {employed the analysis} carried out by \citet{Marel2001AJ....122.1807V} to estimate galactocentric distance in kpc for each identified region. 

\begin{figure*}[htbp]
  \centering
  \subfigure{\includegraphics[width=0.9\columnwidth]{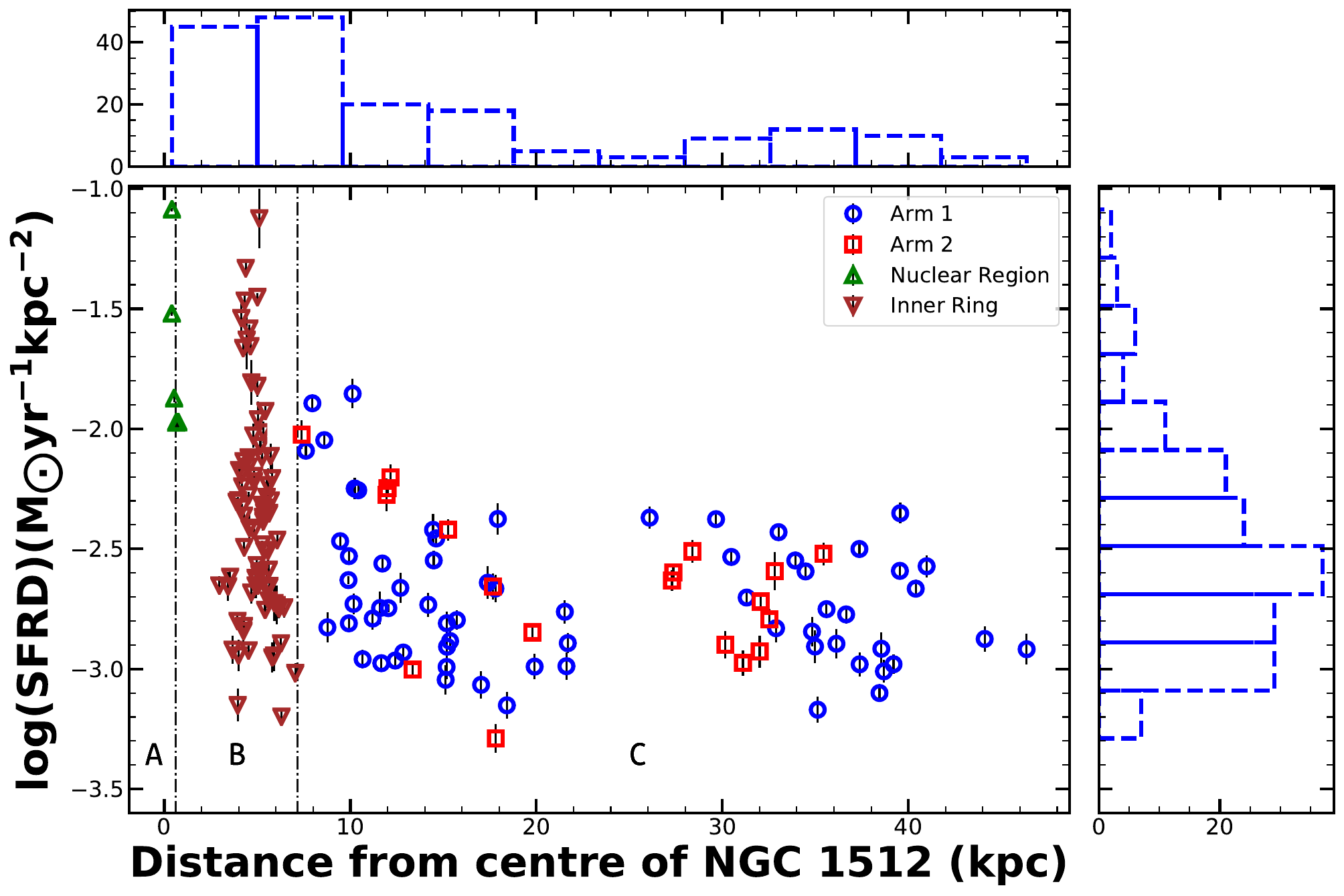}}
  \hfill
  \subfigure{\includegraphics[width=0.9\columnwidth]{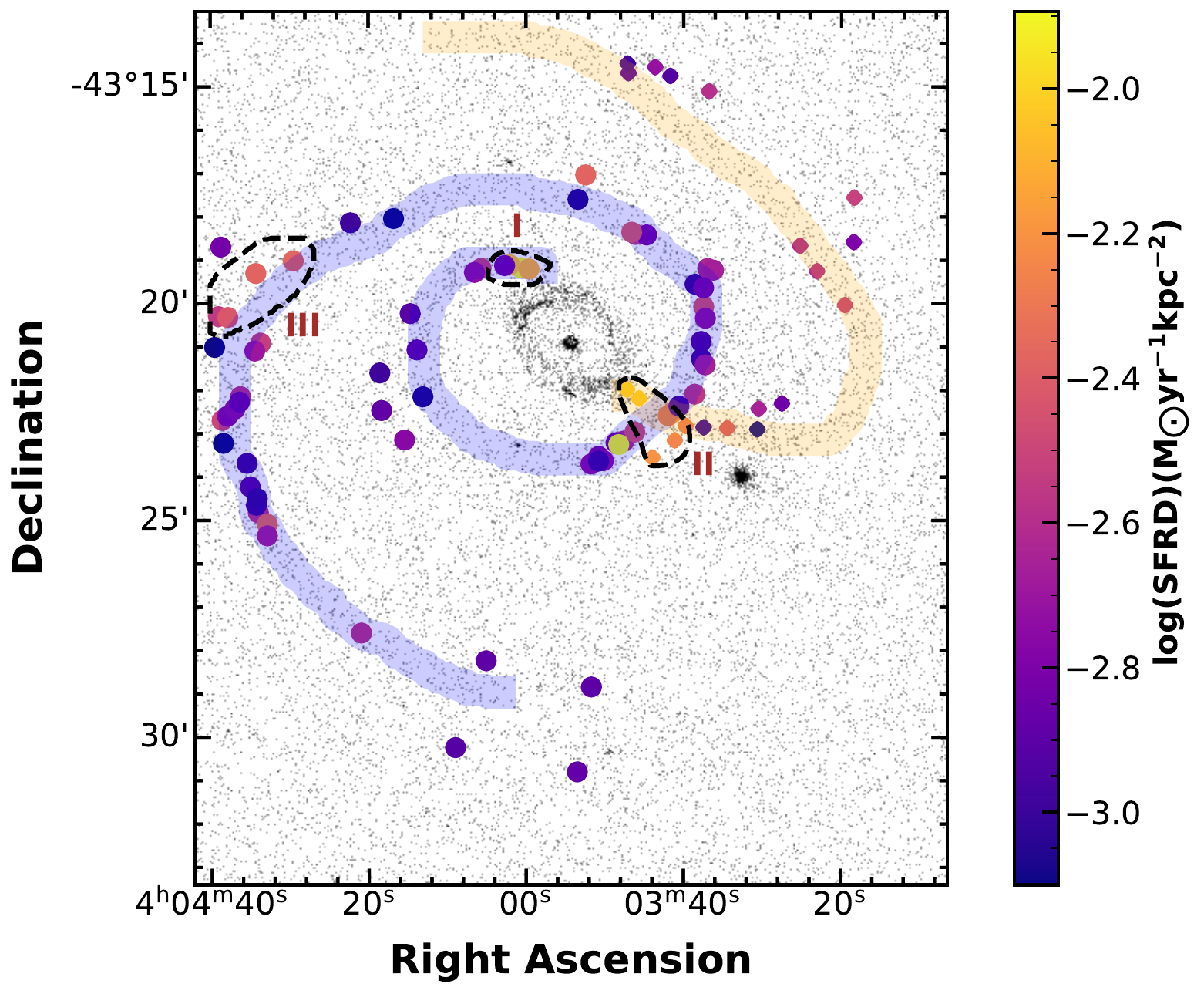}}
  \caption{ (Left) Distribution of SFRD of the regions with respect to the {deprojected} radial distance from the galaxy's centre. {SFRD of a region is calculated by dividing its estimated SFR by the area of the region in kpc$^2$}. It can be observed that the nuclear and inner ring regions of the galaxy (Areas \texttt{A} and \texttt{B}) exhibit higher orders of SFRD (log(SFRD$\mathrm{[M_{\odot}yr^{-1}kpc^{-2}]}) >$ -1.5). On the other hand, the regions of the spiral exhibit lower SFRD (Area \texttt{C}). The histograms parallel to the axes show a distribution of the SFRD and the distance from the centre of NGC 1512 for corresponding regions.
  (Right) The regions identified as part of Arm 1 and Arm 2. The circular symbols represent Arm 1, while the diamond symbols represent Arm 2. {The trajectories of the spiral arms, as identified in Figure \ref{fig:morph} (right panel), are plotted to indicate their positions}. We see an asymmetric distribution of SFRDs. Region \texttt{I} coincides with the origin of Arm 1 and exhibits enhanced star formation (log(SFRD$\mathrm{_{mean}[M_{\odot}yr^{-1}kpc^{-2}]}) \sim$ -2.01). Region \texttt{II} shows local enhancements of star formation, with (log(SFRD$\mathrm{_{mean}[M_{\odot}yr^{-1}kpc^{-2}]}) \sim$ -2.28). Regions in the outer part of Arm 1, denoted by Region \texttt{III}, also show heightened values when compared to other regions of Arm 1, with log(SFRD$\mathrm{_{mean}[M_{\odot}yr^{-1}kpc^{-2}]}) \sim$ -2.36.}
  \label{fig:sfr}
\end{figure*}

We estimated the star formation rate densities (SFRDs) of identified regions by dividing the estimated SFRs with area of each region (in kpc$^2$). Figure \ref{fig:sfr} (left panel) is a radial distribution of the  logarithmic values of SFRD for the identified regions from the galaxy's centre. 
We demarcated different regions of the galaxy as \texttt{A}, \texttt{B} and \texttt{C} on the basis of Figure \ref{fig:morph} (right panel) to understand trends in the spatial distribution of SFRD. \texttt{A} denotes the star-forming regions identified in the nuclear region of the galaxy (represented as green triangles), {\texttt{A} spans an radial distance of $\sim$ 10 \arcsec ($\sim$ 0.6 kpc) from centre of NGC 1512}. \texttt{B} denotes the regions in the inner ring (represented as brown inverted triangles). {\texttt{B} excludes the star-forming regions from \texttt{A} and spans an distance from $\sim$ 10\arcsec - 115\arcsec ($\sim$ 0.6 - 7 kpc)}. \texttt{C} denotes the regions identified as part of the spiral arms. The blue circles denote regions identified as part of Arm 1, and the red squares denote regions identified as part of Arm 2. We note that the nuclear region of the galaxy (\texttt{A}) exhibits heightened rates of star formation. We also observed higher values of SFRD in the inner ring region (\texttt{B}). This shows enhanced star formation occurring in the nuclear region, the circumnuclear ring and the inner ring of the galaxy. {As we move} radially outwards, the SFRD decreases along the spiral arm regions (\texttt{C}), which aligns with the expected distribution of star formation in a typical spiral galaxy \citep{Martel2013MNRAS.431.2560M}. However, some regions, {for instance at radial distances of 12 kpc and 30 kpc}, show small peaks in the SFRD distribution, as is presented in Figure \ref{fig:sfr} (left panel). These localized enhancements in star formation present an interesting radial profile, suggesting the presence of additional factors influencing star formation dynamics in specific regions. 

The observed phenomena in the radial distribution of identified regions can be better understood by studying the local trends in the star formation properties. The study of identified regions in the arms of the galaxy is given in Section \ref{sec:spiralarm} and the study of the inner ring is given in Section \ref{sec:innerring}.
\subsection{Understanding the effect of interaction on the galaxy}\label{sec:spiralarm}

The formation scenario of spiral structures in galaxies has been attributed to different factors that can be secular or environmental in nature. Environmental effects due to interaction, such as tidal effects, may instigate a spiral structure due to the presence of a massive or satellite companion or create distortions in the already existing structures \citep{Toomre1972, Donner1991A&A...252..571D,Struck2011MNRAS.414.2498S}. Thus, studying the spiral arms of the galaxy will help to understand the role of interactions in shaping the spiral arms of the galaxy.  In this context, we study star formation along the spiral arms of NGC 1512. We intend to comprehend and compare the properties of Arm 1 with Arm 2, which were defined in Section \ref{sec:results}. This will help us understand the effect of the main galaxy's interaction with the satellite galaxy. To compare the physical properties of Arm 1 and Arm 2, we performed a two-sample Kolmogorov-Smirnov test (K-S test, \citealp{pratt1981kolmogorov}) on the properties of the star-forming regions to check whether the properties {of Arm 1 and  Arm 2 are significantly different} or not. When the K-S test is performed for the SFRD of the star-forming regions, the {p-value} is {0.27} for the studied population being of the same distribution. When the test is carried out for assessing the distribution of the identified regions' area, the {p-value} is {0.04}. Thus, from the K-S test, we note that there may not be significant differences in the {overall star formation properties} of the identified star-forming regions in the arms. We probe the localized effects of the interaction event on the two spiral arms, if any.

\subsubsection{Studying the star formation properties along the spiral arms}

The interaction episode of NGC 1512 with NGC 1510 has created disruptions in the spiral arms. The interaction has resulted in disruptions to the {regions originating from the SW of the inner ring}, which have formed a tidal bridge towards the satellite galaxy. Furthermore, Arm 1 has undergone visible distortions due to interaction, as is visible in Figure \ref{fig:morph} (left panel). These observations provide valuable insights into the complex interplay between galaxy interactions and the morphology of spiral galaxies. The distortions observed in the spiral arms of NGC 1512 highlight the effects that interaction events can impose on the intricate structure of a galaxy.

In Figure \ref{fig:sfr} (right panel), {we plot} the regions identified in the spiral arms of the galaxy, Arm 1 and  Arm 2. The regions {are colour-coded} based on the intensity of the calculated SFRD. We {observe} three regions in the arms exhibiting higher SFRD values compared to the other regions in the arms (log(SFRD[M$_{\odot}$yr$^{-1}$kpc$^{-2}$]) > -2.4), {which represents the highest $\mathrm{85^{th}}$ percentile in log(SFRD) values}. We consider these three regions of interest, labelled as \texttt{I}, \texttt{II} and \texttt{III} in Figure \ref{fig:sfr} (right panel). These regions have been named in decreasing order of the SFRD values.

We observed heightened SFRD values at the origin of Arm 1, denoted as Region \texttt{I}. The observed enhancement is a possible effect of a combination of orbit crowding, cloud collisions, and gravitational instabilities \citep{Elmegreen2009IAUS..254..289E,Urquhart2021MNRAS.500.3050U}. We estimated a mean log(SFRD[ M$_{\odot}$yr$^{-1}$kpc$^{-2}$]) value of -2.01 $\pm$ 0.02 for this region. Numerical simulations by \citet{Renaud2013MNRAS.436.1836R, Renaud2015MNRAS.454.3299R} investigated the role  of bars in triggering star formation and concluded that the leading edges of the bars favour converging gas flows and large-scale shocks.

Star-forming regions identified in the region of interaction {between the main and satellite galaxy} (Region \texttt{II}) exhibit {higher} SFRD values than other regions associated with Arm 2. This observed enhancement exhibits the important role that the interaction with NGC 1510 plays in shaping the main galaxy. {Region} \texttt{II} has a mean log(SFRD[ M$_{\odot}$yr$^{-1}$kpc$^{-2}$]) value of -2.28  $\pm$ 0.03 in comparison to the mean log(SFRD[ M$_{\odot}$yr$^{-1}$kpc$^{-2}$]) value of {-2.67 $\pm$ 0.04 } for {star-forming regions in Arm 2, excluding Region \texttt{II}}. Thus, the interaction event of NGC 1512 with NGC 1510 has not only disrupted Arm 2 but also might have triggered local bursts of star formation in the vicinity of the interaction. This region is {embedded in} a high-density region in the H\textsc{i} column density map (refer to Figure \ref{fig:morph} (left panel)), thus showing that the interaction event has created a zone of intense star formation in the region of interaction. We also observed that the H\textsc{i} has a non-uniform distribution in this region. The observed enhancement in the star formation regions and H\textsc{i} column density is a possible result of the interaction event between the galaxy pair \citep{Keel1991IAUS..146..243K, Struck1999PhR...321....1S}.

\citet{Teodoro2014A&A...567A..68D} proposed a relation to {estimate the gas accretion rate from minor mergers onto star-forming} galaxies, which is given as:

\begin{equation}
   \dot{\mathrm{M}} \mathrm{_{H\textsc{i}} = \sum_{i=0}^{n} M_{H\textsc{i},i}/T_{0,i}}    
\end{equation}

where the $\mathrm{\dot{M}_{H\textsc{i}}}$ is the mass accretion rate, $\mathrm{M_{HI,i}}$ {denotes the} H\textsc{i} mass of $\mathrm{i^{th}}$ dwarf galaxy and $\mathrm{T_{0,i}}$ {the dynamical time taken for the completion of minor merger event of \texttt{i} dwarf companions with the massive galaxy}. For our case, we consider the case of the dwarf galaxy NGC 1510 as a satellite. Since the galaxy pairs are already extremely close to each other ($\sim$ 14 kpc) and H\textsc{i} envelopes already interacting, we assume that the minor merger will be completed in about a dynamical time of 1 Gyr. We estimated the H\textsc{i} mass of NGC 1510 as $2.84\times 10^7$ M$_{\odot}$ \citep{Barnes2001MNRAS.322..486B}. By substituting the values, we estimate the mass accretion rate of H\textsc{i} gas from the satellite galaxy to be around $2\times 10^{-2}$ M$_{\odot}$yr$^{-1}$. The estimated accretion rate is one-fifth of the global SFR of {NGC 1512 estimated in this study} ($\sim$ 0.11  M$_{\odot}$yr$^{-1}$). This suggests that the {ongoing} accretion event might not be the major contributor to driving the star formation in NGC 1512 \citep{Teodoro2014A&A...567A..68D}. Further analysis might help us understand the other factors influencing star formation in the main galaxy.

Region \texttt{III} belongs to the outer regions of Arm 1. We observe high SFRD values (mean log(SFRD[ M$_{\odot}$yr$^{-1}$kpc$^{-2}$]) value $\sim$ -2.36 $\pm$ 0.04) {in this region}. These regions also exhibit a higher H\textsc{i} column density {compared to other regions identified in the outskirts of Arm 1}. {These are indicators} of ongoing star formation. It is possible that the {enhancement in SFRD and higher H\textsc{i} observed in Region \texttt{III}} is a possible effect of previous interaction events or minor mergers that the main galaxy might have undergone over the last few Gyrs. When a galaxy falls into a larger host galaxy, it can bring along gas that fuels star formation in the outskirts of the host galaxy \citep{Malin1997PASA...14...52M}. {In the past, these remnants} from merger events became diffused and indistinguishable from the other regions of the galaxy \citep{Delgado2009ApJ...692..955M,Delgado2010AJ....140..962M}. {Therefore, we realize that past interactions of NGC 1512 are a possible reason for the observed morphology of spiral arms and local enhancements of star formation on the outskirts of the spiral arms.}

Based on the observations of localised enhancements of SFRD and higher H\textsc{i} column density in NGC 1512, {we suggest} that past and ongoing interaction events may have instigated local bursts of star formation and contributed to the shaping of the galaxy. The enhanced SFRD and higher H\textsc{i} column density observed in these regions are potentially a result of interaction episodes: with the satellite galaxy NGC 1510 in Region \texttt{II} and older interaction episodes in the outskirts of Arm 1. Our observations show a positive correlation between the regions exhibiting higher H\textsc{i} column density and enhancements in SFRD. Therefore, the observed enhancement in star formation and the morphology of the spiral arms in NGC 1512 suggest that interaction events have played a significant role in the {evolutionary history} of the galaxy. The effects of these interactions may have been long-lasting, with the remnants of past mergers and interactions becoming diffused over time and integrated into the overall structure of the galaxy. 

\subsection{Star formation in the inner ring} \label{sec:innerring}

\begin{figure}
\centering
{\includegraphics[ width=1\columnwidth]{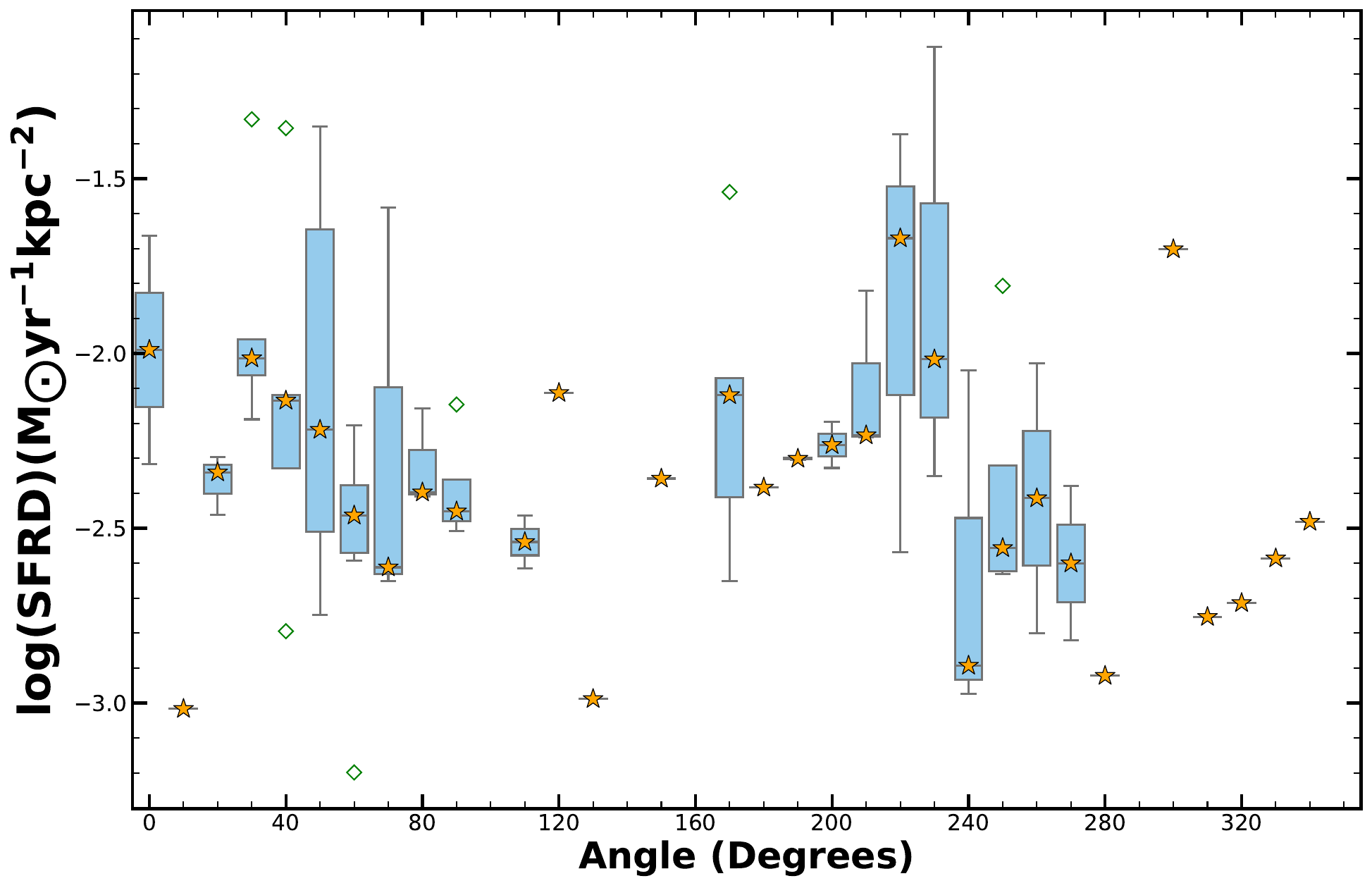}}
\caption{We examine the distribution of SFRD in the identified regions of the inner ring of the galaxy (region shown in the inset of Figure \ref{fig:composite}) based on their angles relative to {North in a boxplot}. {The box represents the interquartile range (IQR) of the observed SFRD, the orange stars indicate the median of SFRD, and the whiskers depict data points within the range of 1.5 times the IQR. Outliers, depicted as green diamonds beyond the whisker limits, are values that fall outside this range. Individual hyphen markers at some places indicate that only a single star-forming region subtends the corresponding angle}. We observe a peaking of the SFRD at the area of the inner ring in conjunction with the bar ends.}
\label{fig:innerring}
\end{figure}

While the interaction with its satellite galaxy has had {significant effects} on the morphology as well as the star formation properties of NGC 1512, the presence of a strong bar in the galaxy ({bar strength of $\mathrm{Q_b} \sim$ 0.27; \citealp{Buta2005AJ....130..506B,Buta2006AJ....132.1859B}}), {can influence its evolution}. 

\citet{Schwarz1981ApJ...247...77S, Schwarz1984MNRAS.209...93S} discussed the formation of rings in a galaxy as a result of the secular evolution in galaxies due to bar-driven gas flow. The presence of a bar can funnel the cold gas inside the co-rotation radius to the nuclear region, leading to a dearth of cold gas in the co-rotation radius region \citep{Haywood2016A&A...589A..66H,George2019A&A...621L...4G,george2020A&A...644A..79G}.

Previous studies have been unable to resolve the star-forming regions in the inner ring of NGC 1512. However, owing to the superior resolving power of UVIT, we have been able to resolve individual star-forming regions, allowing us to study the properties of the regions in detail. The inner ring {holds the promise of new discoveries} as it encircles the bar at {3\arcmin} from the galactic centre and has been largely unexplored. Our main goal for this analysis is to study the physical properties of {the identified regions} in the inner ring and, thus, understand the effect of the galactic bar on the inner ring.

We observed a concentration of star-forming regions in the inner ring region near the major axis of the bar, as is visible in Figure \ref{fig:morph} (left panel). We note from Figure \ref{fig:sfr} (left panel) that the {inner ring shows} the highest values for SFRD ($\mathrm{log(SFRD_{mean} [M_{\odot}yr^{-1}kpc^{-2}]) \sim -1.7}$). The cause for such heightened values of SFRD needs to be understood. We used the deprojected coordinate values for each region such that the major axis of the galaxy coincides with the X-axis. We then estimated the angle subtended by each region with respect to {North} in the counter-clockwise direction, with 0$^\circ$ denoting the positive Y-axis. Figure \ref{fig:innerring} represents the distribution of SFRD along the angle subtended by the regions with the inner ring of the galaxy. We observe in Figure \ref{fig:innerring} that the regions lying in {the region of the conjunction} between the end of the bar and the inner ring, near 45$^\circ$ and 225$^\circ$ show heightened star formation. There is a clustering of star-forming regions around the major axis of the bar which shows heightened star-formation activity.

{The formation of inner ring} in galaxies have been explained due to gas being caught at the Ultraharmonic Resonance (UHR) position of the galaxy, preventing it from being funnelled by the bar towards the galactic centre \citep{Schwarz1984MNRAS.209...93S,Lord_BARCROWDING1991ApJ...381..130L,Buta2004AJ....127.1982B,Byrd2006AJ....131.1377B}. We understand that due to further compression and shocks in the trapped gas, leading to a possible crowding effect, {star formation can be} enhanced in this region. This is supported by the observed trends in SFRD for the inner ring, with heightened star formation activity around the major axis of the bar.

We studied the star formation activity {inside the region of the bar (as identified in the left panel of Figure \ref{fig:morph})} of the galaxy and found a distinct lack of UV emission {in this region}. This suggests that {there is no ongoing or recent star formation occurring in this region}. To further understand the possible reason behind the absence of star formation {inside the bar region}, we overlaid the H\textsc{i} column density over the UVIT image {in Figure \ref{fig:morph} (left panel)}. Our analysis showed a lack of H\textsc{i} distribution in this region when compared to the surrounding regions. Such a distribution hints at a possible variation in the distribution of neutral hydrogen gas and a potential deviation from typical gas dynamics in the observed region. This indicates that the H\textsc{i} gas in the galaxy is being redistributed, and its distribution {inside the bar region} is affected by the presence of the galactic bar (eg.: \citealp{George2019A&A...621L...4G,george2020A&A...644A..79G,Saha2021JApA...42...59S}). The identified region is also devoid of CO(2-1) gas \citep{Lee2023ApJ...944L..17L}. These observations highlight the importance of galactic bars on star formation and H\textsc{i} gas distribution in galaxies. It is crucial to understand how the presence of galactic bars affects the evolution of galaxies and how it contributes to the overall distribution of H\textsc{i} gas and star formation in a galaxy.

Based on our analysis, it can be inferred that the galactic bar plays a significant role in driving the secular evolution of NGC 1512. This is in addition to the effects of past and ongoing interaction events that may also be contributing to the observed properties of the galaxy. We find that the galactic bar plays a critical role in redistributing H\textsc{i} gas in the inner regions of the galaxy and the subsequent quenching of star formation {inside the bar radius}.
The redistribution of the H\textsc{i} gas in the central regions of the galaxy is likely driven by the non-axisymmetric gravitational potential created by the bar. {This is the first study to understand the combined effects of the galactic bar and the interaction with the satellite galaxy on the evolution of NGC 1512}.

Our findings demonstrate that the evolution of NGC 1512 is influenced by a combination of secular and environmental factors. The interplay between these effects is crucial in shaping the observed properties and {evolution of NGC 1512}.

\subsection{Star formation in {the} satellite galaxy, NGC 1510} \label{sec:1510sfr}

The satellite galaxy of NGC 1510 is classified as a Blue Compact Dwarf (BCD) galaxy. The satellite galaxy is situated at a distance of $\sim$ $\mathrm{5\arcmin}$ ($\mathrm{\sim}$ {18.3 kpc}) from the main galaxy. We observed the satellite galaxy to be embedded in a high H\textsc{i} gas density region ($\mathrm{>0.45\times10^{21} cm^{-2}}$). We identified three star-forming regions in the galaxy with values of $\mathrm{log(SFRD[M_{\odot}yr^{-1}kpc^{-2}])}$ ranging from -0.71 to -2.65. We {conclude} that the satellite galaxy is experiencing enhanced levels of star formation. \citet{Hawarden1979A&A....76..230H} suggested the observed blue colour of the satellite galaxy was a result of the {high H\textsc{i} column density region it contains}. The observed high H\textsc{i} column density region in which the satellite galaxy is embedded may be due to {gas-capture} by the satellite galaxy from the main galaxy \citep{Hawarden1979A&A....76..230H,Gallagher2005nfcd.conf..151G}.

\section{Summary} \label{summary}

We conducted this study to investigate the impact of interaction events and the galactic bar {on the evolution of the interacting} galaxy pair NGC 1512/1510. The aim of this study is to provide insight into the driving factors for the observed morphological and star-forming properties of NGC 1512 and to elucidate the mechanisms that govern galactic evolution in the context of interaction and secular processes. A summary of the main results obtained from our study is given below.
\begin{itemize}
    \item We used the UVIT FUV and NUV observations to identify and characterize the star-forming regions in the galaxy pair.
    
    \item We identified 175 star-forming regions in the UVIT FUV image of NGC 1512 {using the ProFound} package. The calculated magnitudes were extinction corrected.
    \item We identified regions which exhibited enhanced SFRD. We observed that in addition to the enhancement of star-formation near the origin of Arm 1, enhancement is observed in the region of interaction between the galaxy pair and {on the outskirts of Arm 1}. {These star-forming regions have high H\textsc{i} density}. 
    \item The observed enhancements in SFRD and H\textsc{i} density in the interaction regions between the galaxy pair and the outskirts of Arm 1 indicate that the {evolution of NGC 1512} is intimately linked to its interaction history.
   
   \item We observed heightened star formation activity in the inner ring near the ends of the galactic bar. The conspicuous lack of both UV and H\textsc{i} emissions detected within {the bar radius} points towards the redistribution of gas in the central regions of the galaxy. This action is seen to force the gas to migrate inwards towards the central regions of the galaxy, causing a depletion of gas supply and can eventually lead to a suppression of star formation activity.
   
    \item Our study found enhanced star formation in NGC 1510. {This is consistent with} the high H\textsc{i} column density region ( $> 0.45 \times 10^{21}$cm$^{-2}$) {within the galaxy}.
    
\end{itemize}
We conclude that the galaxy NGC 1512 owes its {present state to different evolutionary factors}, both environmental and secular. To further quantify the effect of each of these factors in galactic evolution, numerical simulations will be crucial. These simulations can help us understand the details of the gas dynamics and star formation history of the galaxy and provide more insight on how each factor has influenced the evolution of NGC 1512.

\section*{Acknowledgements}

We thank the anonymous referee for the valuable comments and suggestions that have improved the quality of the paper. We acknowledge the financial support from the Indian Space Research Organisation (ISRO) under the \textit{AstroSat} archival data utilization program (No. DS-2B-13013(2)/6/2019) and the Department of Science and Technology (DST) for the INSPIRE FELLOWSHIP (IF180855). We thank our colleagues Shridharan Baskaran, Akhil Krishna R, Cysil Tom Baby and Belinda Damian for their valuable comments on the manuscript. This publication uses the data from the UVIT, which is part of the \textit{AstroSat} mission of the ISRO, archived at the Indian Space Science Data Centre (ISSDC). We gratefully thank all the individuals involved in the various teams for supporting the project from the early stages of the design to launch and observations with it in orbit. We thank Joseph Postma, University of Calgary, for his consistent support during the process of UVIT data reduction. We thank the Center for Research, CHRIST (Deemed to be University) for all their support during the course of this work. This research has used the NASA/IPAC Extragalactic Database (NED), funded by the National Aeronautics and Space Administration and operated by the California Institute of Technology.

\bibliographystyle{aa}
\bibliography{ref}

\begin{thebibliography}{85}
\expandafter\ifx\csname natexlab\endcsname\relax\def\natexlab#1{#1}\fi

\bibitem[{{Astropy Collaboration} {et~al.}(2013){Astropy Collaboration},
  {Robitaille}, {Tollerud}, {Greenfield}, {Droettboom}, {Bray}, {Aldcroft},
  {Davis}, {Ginsburg}, {Price-Whelan}, {Kerzendorf}, {Conley}, {Crighton},
  {Barbary}, {Muna}, {Ferguson}, {Grollier}, {Parikh}, {Nair}, {Unther},
  {Deil}, {Woillez}, {Conseil}, {Kramer}, {Turner}, {Singer}, {Fox}, {Weaver},
  {Zabalza}, {Edwards}, {Azalee Bostroem}, {Burke}, {Casey}, {Crawford},
  {Dencheva}, {Ely}, {Jenness}, {Labrie}, {Lim}, {Pierfederici}, {Pontzen},
  {Ptak}, {Refsdal}, {Servillat}, \& {Streicher}}]{Astropy2013A&A...558A..33A}
{Astropy Collaboration}, {Robitaille}, T.~P., {Tollerud}, E.~J., {et~al.} 2013,
  \aap, 558, A33

\bibitem[{{Barnes} {et~al.}(2001){Barnes}, {Staveley-Smith}, {de Blok},
  {Oosterloo}, {Stewart}, {Wright}, {Banks}, {Bhathal}, {Boyce}, {Calabretta},
  {Disney}, {Drinkwater}, {Ekers}, {Freeman}, {Gibson}, {Green}, {Haynes}, {te
  Lintel Hekkert}, {Henning}, {Jerjen}, {Juraszek}, {Kesteven}, {Kilborn},
  {Knezek}, {Koribalski}, {Kraan-Korteweg}, {Malin}, {Marquarding}, {Minchin},
  {Mould}, {Price}, {Putman}, {Ryder}, {Sadler}, {Schr{\"o}der}, {Stootman},
  {Webster}, {Wilson}, \& {Ye}}]{Barnes2001MNRAS.322..486B}
{Barnes}, D.~G., {Staveley-Smith}, L., {de Blok}, W.~J.~G., {et~al.} 2001,
  \mnras, 322, 486

\bibitem[{{Bevington} \& {Robinson}(1992)}]{Bevington1992drea.book.....B}
{Bevington}, P.~R. \& {Robinson}, D.~K. 1992, {Data reduction and error
  analysis for the physical sciences}

\bibitem[{{Borgohain} {et~al.}(2022){Borgohain}, {Saha}, {Elmegreen}, {Gogoi},
  {Combes}, \& {Tandon}}]{Borgohain2022Natur.607..459B}
{Borgohain}, A., {Saha}, K., {Elmegreen}, B., {et~al.} 2022, \nat, 607, 459

\bibitem[{{Bresolin} {et~al.}(2012){Bresolin}, {Kennicutt}, \&
  {Ryan-Weber}}]{Bresolin2012}
{Bresolin}, F., {Kennicutt}, R.~C., \& {Ryan-Weber}, E. 2012, \apj, 750, 122

\bibitem[{{Buta} {et~al.}(2006){Buta}, {Laurikainen}, {Salo}, {Block}, \&
  {Knapen}}]{Buta2006AJ....132.1859B}
{Buta}, R., {Laurikainen}, E., {Salo}, H., {Block}, D.~L., \& {Knapen}, J.~H.
  2006, \aj, 132, 1859

\bibitem[{{Buta} {et~al.}(2005){Buta}, {Vasylyev}, {Salo}, \&
  {Laurikainen}}]{Buta2005AJ....130..506B}
{Buta}, R., {Vasylyev}, S., {Salo}, H., \& {Laurikainen}, E. 2005, \aj, 130,
  506

\bibitem[{{Buta} {et~al.}(2004){Buta}, {Byrd}, \&
  {Freeman}}]{Buta2004AJ....127.1982B}
{Buta}, R.~J., {Byrd}, G.~G., \& {Freeman}, T. 2004, \aj, 127, 1982

\bibitem[{{Byrd} {et~al.}(2006){Byrd}, {Freeman}, \&
  {Buta}}]{Byrd2006AJ....131.1377B}
{Byrd}, G.~G., {Freeman}, T., \& {Buta}, R.~J. 2006, \aj, 131, 1377

\bibitem[{{Cardelli} {et~al.}(1989){Cardelli}, {Clayton}, \&
  {Mathis}}]{Cardelli1989ApJ...345..245C}
{Cardelli}, J.~A., {Clayton}, G.~C., \& {Mathis}, J.~S. 1989, \apj, 345, 245

\bibitem[{{Chakrabarti} {et~al.}(2011){Chakrabarti}, {Bigiel}, {Chang}, \&
  {Blitz}}]{Chakrabarti2011ApJ...743...35C}
{Chakrabarti}, S., {Bigiel}, F., {Chang}, P., \& {Blitz}, L. 2011, \apj, 743,
  35

\bibitem[{{de Vaucouleurs} {et~al.}(1991){de Vaucouleurs}, {de Vaucouleurs},
  {Corwin}, {Buta}, {Paturel}, \& {Fouque}}]{deVaucouleurs1991}
{de Vaucouleurs}, G., {de Vaucouleurs}, A., {Corwin}, Herold~G., J., {et~al.}
  1991, {Third Reference Catalogue of Bright Galaxies}

\bibitem[{{Devaraj} {et~al.}(2023){Devaraj}, {Joseph}, {Stalin}, {Tandon}, \&
  {Ghosh}}]{Devaraj2023ApJ...946...65D}
{Devaraj}, A., {Joseph}, P., {Stalin}, C.~S., {Tandon}, S.~N., \& {Ghosh},
  S.~K. 2023, \apj, 946, 65

\bibitem[{{Di Teodoro} \& {Fraternali}(2014)}]{Teodoro2014A&A...567A..68D}
{Di Teodoro}, E.~M. \& {Fraternali}, F. 2014, \aap, 567, A68

\bibitem[{{Donner} {et~al.}(1991){Donner}, {Engstrom}, \&
  {Sundelius}}]{Donner1991A&A...252..571D}
{Donner}, K.~J., {Engstrom}, S., \& {Sundelius}, B. 1991, \aap, 252, 571

\bibitem[{{Doyle} \& {Drinkwater}(2006)}]{Doyle2006MNRAS.372..977D}
{Doyle}, M.~T. \& {Drinkwater}, M.~J. 2006, \mnras, 372, 977

\bibitem[{{Ducci} {et~al.}(2014){Ducci}, {Kavanagh}, {Sasaki}, \&
  {Koribalski}}]{Ducci2014A&A...566A.115D}
{Ducci}, L., {Kavanagh}, P.~J., {Sasaki}, M., \& {Koribalski}, B.~S. 2014,
  \aap, 566, A115

\bibitem[{{Elmegreen}(2009)}]{Elmegreen2009IAUS..254..289E}
{Elmegreen}, B.~G. 2009, in The Galaxy Disk in Cosmological Context, ed.
  J.~{Andersen}, {Nordstr{\"o}ara}, B.~{m}, \& J.~{Bland-Hawthorn}, Vol. 254,
  289--300

\bibitem[{{Gaia Collaboration} {et~al.}(2022){Gaia Collaboration}, {Vallenari},
  {Brown}, {Prusti}, {de Bruijne}, {Arenou}, {Babusiaux}, {Biermann},
  {Creevey}, {Ducourant}, {Evans}, {Eyer}, {Guerra}, {Hutton}, {Jordi},
  {Klioner}, {Lammers}, {Lindegren}, {Luri}, {Mignard}, {Panem}, {Pourbaix},
  {Randich}, {Sartoretti}, {Soubiran}, {Tanga}, {Walton}, {Bailer-Jones},
  {Bastian}, {Drimmel}, {Jansen}, {Katz}, {Lattanzi}, {van Leeuwen}, {Bakker},
  {Cacciari}, {Casta{\~n}eda}, {De Angeli}, {Fabricius}, {Fouesneau},
  {Fr{\'e}mat}, {Galluccio}, {Guerrier}, {Heiter}, {Masana}, {Messineo},
  {Mowlavi}, {Nicolas}, {Nienartowicz}, {Pailler}, {Panuzzo}, {Riclet}, {Roux},
  {Seabroke}, {Sordo{\o}rcit}, {Th{\'e}venin}, {Gracia-Abril}, {Portell},
  {Teyssier}, {Altmann}, {Andrae}, {Audard}, {Bellas-Velidis}, {Benson},
  {Berthier}, {Blomme}, {Burgess}, {Busonero}, {Busso}, {C{\'a}novas}, {Carry},
  {Cellino}, {Cheek}, {Clementini}, {Damerdji}, {Davidson}, {de Teodoro},
  {Nu{\~n}ez Campos}, {Delchambre}, {Dell'Oro}, {Esquej},
  {Fern{\'a}ndez-Hern{\'a}ndez}, {Fraile}, {Garabato}, {Garc{\'\i}a-Lario},
  {Gosset}, {Haigron}, {Halbwachs}, {Hambly}, {Harrison}, {Hern{\'a}ndez},
  {Hestroffer}, {Hodgkin}, {Holl}, {Jan{\ss}en}, {Jevardat de Fombelle},
  {Jordan}, {Krone-Martins}, {Lanzafame}, {L{\"o}ffler}, {Marchal}, {Marrese},
  {Moitinho}, {Muinonen}, {Osborne}, {Pancino}, {Pauwels}, {Recio-Blanco},
  {Reyl{\'e}}, {Riello}, {Rimoldini}, {Roegiers}, {Rybizki}, {Sarro}, {Siopis},
  {Smith}, {Sozzetti}, {Utrilla}, {van Leeuwen}, {Abbas}, {{\'A}brah{\'a}m},
  {Abreu Aramburu}, {Aerts}, {Aguado}, {Ajaj}, {Aldea-Montero}, {Altavilla},
  {{\'A}lvarez}, {Alves}, {Anders}, {Anderson}, {Anglada Varela}, {Antoja},
  {Baines}, {Baker}, {Balaguer-N{\'u}{\~n}ez}, {Balbinot}, {Balog}, {Barache},
  {Barbato}, {Barros}, {Barstow}, {Bartolom{\'e}}, {Bassilana}, {Bauchet},
  {Becciani}, {Bellazzini}, {Berihuete}, {Bernet}, {Bertone}, {Bianchi},
  {Binnenfeld}, {Blanco-Cuaresma}, {Blazere}, {Boch}, {Bombrun}, {Bossini},
  {Bouquillon}, {Bragaglia}, {Bramante}, {Breedt}, {Bressan}, {Brouillet},
  {Brugaletta}, {Bucciarelli}, {Burlacu}, {Butkevich}, {Buzzi}, {Caffau},
  {Cancelliere}, {Cantat-Gaudin}, {Carballo}, {Carlucci}, {Carnerero},
  {Carrasco}, {Casamiquela}, {Castellani}, {Castro-Ginard}, {Chaoul},
  {Charlot}, {Chemin}, {Chiaramida}, {Chiavassa}, {Chornay}, {Comoretto},
  {Contursi}, {Cooper}, {Cornez}, {Cowell}, {Crifo}, {Cropper}, {Crosta},
  {Crowley}, {Dafonte}, {Dapergolas}, {David}, {David}, {de Laverny}, {De
  Luise}, {De March}, {De Ridder}, {de Souza}, {de Torres}, {del Peloso}, {del
  Pozo}, {Delbo}, {Delgado}, {Delisle}, {Demouchy}, {Dharmawardena}, {Di
  Matteo}, {Diakite}, {Diener}, {Distefano}, {Dolding}, {Edvardsson}, {Enke},
  {Fabre}, {Fabrizio}, {Faigler}, {Fedorets}, {Fernique}, {Fienga}, {Figueras},
  {Fournier}, {Fouron}, {Fragkoudi}, {Gai}, {Garcia-Gutierrez},
  {Garcia-Reinaldos}, {Garc{\'\i}a-Torres}, {Garofalo}, {Gavel}, {Gavras},
  {Gerlach}, {Geyer}, {Giacobbe}, {Gilmore}, {Girona}, {Giuffrida}, {Gomel},
  {Gomez}, {Gonz{\'a}lez-N{\'u}{\~n}ez}, {Gonz{\'a}lez-Santamar{\'\i}a},
  {Gonz{\'a}lez-Vidal}, {Granvik}, {Guillout}, {Guiraud},
  {Guti{\'e}rrez-S{\'a}nchez}, {Guy}, {Hatzidimitriou}, {Hauser}, {Haywood},
  {Helmer}, {Helmi}, {Sarmiento}, {Hidalgo}, {Hilger}, {H{\l}adczuk}, {Hobbs},
  {Holland}, {Huckle}, {Jardine}, {Jasniewicz}, {Jean-Antoine Piccolo},
  {Jim{\'e}nez-Arranz}, {Jorissen}, {Juaristi Campillo}, {Julbe}, {Karbevska},
  {Kervella}, {Khanna}, {Kontizas}, {Kordopatis}, {Korn}, {K{\'o}sp{\'a}l},
  {Kostrzewa-Rutkowska}, {Kruszy{\'n}ska}, {Kun}, {Laizeau}, {Lambert},
  {Lanza}, {Lasne}, {Le Campion}, {Lebreton}, {Lebzelter}, {Leccia}, {Leclerc},
  {Lecoeur-Taibi}, {Liao}, {Licata}, {Lindstr{\o}m}, {Lister}, {Livanou},
  {Lobel}, {Lorca}, {Loup}, {Madrero Pardo}, {Magdaleno Romeo}, {Managau},
  {Mann}, {Manteiga}, {Marchant}, {Marconi}, {Marcos}, {Marcos Santos},
  {Mar{\'\i}n Pina}, {Marinoni}, {Marocco}, {Marshall}, {Polo},
  {Mart{\'\i}n-Fleitas}, {Marton}, {Mary}, {Masip}, {Massari},
  {Mastrobuono-Battisti}, {Mazeh}, {McMillan}, {Messina}, {Michalik}, {Millar},
  {Mints}, {Molina}, {Molinaro}, {Moln{\'a}r}, {Monari}, {Mongui{\'o}},
  {Montegriffo}, {Montero}, {Mor}, {Mora}, {Morbidelli}, {Morel}, {Morris},
  {Muraveva}, {Murphy}, {Musella}, {Nagy}, {Noval}, {Oca{\~n}a}, {Ogden},
  {Ordenovic}, {Osinde}, {Pagani}, {Pagano}, {Palaversa}, {Palicio},
  {Pallas-Quintela}, {Panahi}, {Payne-Wardenaar}, {Pe{\~n}alosa Esteller},
  {Penttil{\"a}}, {Pichon}, {Piersimoni}, {Pineau}, {Plachy}, {Plum}, {Poggio},
  {Pr{\v{s}}a}, {Pulone}, {Racero}, {Ragaini}, {Rainer}, {Raiteri}, {Rambaux},
  {Ramos}, {Ramos-Lerate}, {Re Fiorentin}, {Regibo}, {Richards}, {Rios Diaz},
  {Ripepi}, {Riva}, {Rix}, {Rixon}, {Robichon}, {Robin}, {Robin}, {Roelens},
  {Rogues}, {Rohrbasser}, {Romero-G{\'o}mez}, {Rowell}, {Royer}, {Ruz Mieres},
  {Rybicki}, {Sadowski}, {S{\'a}ez N{\'u}{\~n}ez}, {Sagrist{\`a} Sell{\'e}s},
  {Sahlmann}, {Salguero}, {Samaras}, {Sanchez Gimenez}, {Sanna},
  {Santove{\~n}a}, {Sarasso}, {Schultheis}, {Sciacca}, {Segol}, {Segovia},
  {S{\'e}gransan}, {Semeux}, {Shahaf}, {Siddiqui}, {Siebert}, {Siltala},
  {Silvelo}, {Slezak}, {Slezak}, {Smart}, {Snaith}, {Solano}, {Solitro},
  {Souami}, {Souchay}, {Spagna}, {Spina}, {Spoto}, {Steele},
  {Steidelm{\"u}ller}, {Stephenson}, {S{\"u}veges}, {Surdej}, {Szabados},
  {Szegedi-Elek}, {Taris}, {Taylo}, {Teixeira}, {Tolomei}, {Tonello}, {Torra},
  {Torra}, {Torralba Elipe}, {Trabucchi}, {Tsounis}, {Turon}, {Ulla}, {Unger},
  {Vaillant}, {van Dillen}, {van Reeven}, {Vanel}, {Vecchiato}, {Viala},
  {Vicente}, {Voutsinas}, {Weiler}, {Wevers}, {Wyrzykowski}, {Yoldas}, {Yvard},
  {Zhao}, {Zorec}, {Zucker}, \& {Zwitter}}]{Gaia2022arXiv220800211G}
{Gaia Collaboration}, {Vallenari}, A., {Brown}, A.~G.~A., {et~al.} 2022, arXiv
  e-prints, arXiv:2208.00211

\bibitem[{{Gallagher} {et~al.}(2005){Gallagher}, {Grebel}, \&
  {Smith}}]{Gallagher2005nfcd.conf..151G}
{Gallagher}, J.~S., {Grebel}, E.~K., \& {Smith}, L.~J. 2005, in IAU Colloq.
  198: Near-fields cosmology with dwarf elliptical galaxies, ed. H.~{Jerjen} \&
  B.~{Binggeli}, 151--155

\bibitem[{{George} {et~al.}(2019){George}, {Joseph}, {Mondal}, {Subramanian},
  {Subramaniam}, \& {Paul}}]{George2019A&A...621L...4G}
{George}, K., {Joseph}, P., {Mondal}, C., {et~al.} 2019, \aap, 621, L4

\bibitem[{{George} {et~al.}(2020){George}, {Joseph}, {Mondal}, {Subramanian},
  {Subramaniam}, \& {Paul}}]{george2020A&A...644A..79G}
{George}, K., {Joseph}, P., {Mondal}, C., {et~al.} 2020, \aap, 644, A79

\bibitem[{{Girish} {et~al.}(2017){Girish}, {Tandon}, {Sriram}, {Kumar}, \&
  {Postma}}]{Girish2017ExA....43...59G}
{Girish}, V., {Tandon}, S.~N., {Sriram}, S., {Kumar}, A., \& {Postma}, J. 2017,
  Experimental Astronomy, 43, 59

\bibitem[{{Hawarden} {et~al.}(1979){Hawarden}, {van Woerden}, {Mebold}, {Goss},
  \& {Peterson}}]{Hawarden1979A&A....76..230H}
{Hawarden}, T.~G., {van Woerden}, H., {Mebold}, U., {Goss}, W.~M., \&
  {Peterson}, B.~A. 1979, \aap, 76, 230

\bibitem[{{Haywood} {et~al.}(2016){Haywood}, {Lehnert}, {Di Matteo}, {Snaith},
  {Schultheis}, {Katz}, \& {G{\'o}mez}}]{Haywood2016A&A...589A..66H}
{Haywood}, M., {Lehnert}, M.~D., {Di Matteo}, P., {et~al.} 2016, \aap, 589, A66

\bibitem[{{Healy} {et~al.}(2021){Healy}, {Deb}, {Verheijen}, {Blyth}, {Serra},
  {Ramatsoku}, \& {Vulcani}}]{Healy2021A&A...654A.173H}
{Healy}, J., {Deb}, T., {Verheijen}, M.~A.~W., {et~al.} 2021, \aap, 654, A173

\bibitem[{Jonas(2018)}]{Jonas:2018Jr}
Jonas, J. 2018, in Proceedings of MeerKAT Science: On the Pathway to the SKA
  {\textemdash} PoS(MeerKAT2016), Vol. 277, 001

\bibitem[{{Jonas} \& {MeerKAT Team}(2016)}]{Jonas2016mks..confE...1J}
{Jonas}, J. \& {MeerKAT Team}. 2016, in MeerKAT Science: On the Pathway to the
  SKA, 1

\bibitem[{{J{\'o}zsa} {et~al.}(2020){J{\'o}zsa}, {White}, {Thorat}, {Smirnov},
  {Serra}, {Ramatsoku}, {Ramaila}, {Perkins}, {Maccagni}, {Makhathini},
  {Moln{\'a}r}, {Kamphuis}, {Kleiner}, {Hugo}, {de Blok}, \&
  {Andati}}]{Jozsa2020ASPC..527..635J}
{J{\'o}zsa}, G.~I.~G., {White}, S.~V., {Thorat}, K., {et~al.} 2020, in
  Astronomical Society of the Pacific Conference Series, Vol. 527, Astronomical
  Data Analysis Software and Systems XXIX, ed. R.~{Pizzo}, E.~R. {Deul}, J.~D.
  {Mol}, J.~{de Plaa}, \& H.~{Verkouter}, 635

\bibitem[{{Karachentsev} \& {Kaisina}(2013)}]{Kaisina2013}
{Karachentsev}, I.~D. \& {Kaisina}, E.~I. 2013, \aj, 146, 46

\bibitem[{{Karthick} {et~al.}(2014){Karthick}, {L{\'o}pez-S{\'a}nchez}, {Sahu},
  {Sanwal}, \& {Bisht}}]{Karthick2014MNRAS.439..157K}
{Karthick}, M.~C., {L{\'o}pez-S{\'a}nchez}, {\'A}.~R., {Sahu}, D.~K., {Sanwal},
  B.~B., \& {Bisht}, S. 2014, \mnras, 439, 157

\bibitem[{{Keel}(1991)}]{Keel1991IAUS..146..243K}
{Keel}, W.~C. 1991, in Dynamics of Galaxies and Their Molecular Cloud
  Distributions, ed. F.~{Combes} \& F.~{Casoli}, Vol. 146, 243

\bibitem[{{Kennicutt}(1998)}]{Kennicutt1998}
{Kennicutt}, Robert~C., J. 1998, \araa, 36, 189

\bibitem[{{Kennicutt} \& {Evans}(2012)}]{Kennicutt2012ARA&A..50..531K}
{Kennicutt}, R.~C. \& {Evans}, N.~J. 2012, \araa, 50, 531

\bibitem[{{Kinman}(1978)}]{Kinman_arms1978AJ.....83..764K}
{Kinman}, T.~D. 1978, \aj, 83, 764

\bibitem[{{Komatsu} {et~al.}(2011){Komatsu}, {Smith}, {Dunkley}, {Bennett},
  {Gold}, {Hinshaw}, {Jarosik}, {Larson}, {Nolta}, {Page}, {Spergel},
  {Halpern}, {Hill}, {Kogut}, {Limon}, {Meyer}, {Odegard}, {Tucker}, {Weiland},
  {Wollack}, \& {Wright}}]{Komatsu2011ApJS..192...18K}
{Komatsu}, E., {Smith}, K.~M., {Dunkley}, J., {et~al.} 2011, \apjs, 192, 18

\bibitem[{{Koribalski} \& {L{\'o}pez-S{\'a}nchez}(2009)}]{Koribalski_1512_2009}
{Koribalski}, B.~S. \& {L{\'o}pez-S{\'a}nchez}, {\'A}.~R. 2009, \mnras, 400,
  1749

\bibitem[{{Koribalski} {et~al.}(2004){Koribalski}, {Staveley-Smith}, {Kilborn},
  {Ryder}, {Kraan-Korteweg}, {Ryan-Weber}, {Ekers}, {Jerjen}, {Henning},
  {Putman}, {Zwaan}, {de Blok}, {Calabretta}, {Disney}, {Minchin}, {Bhathal},
  {Boyce}, {Drinkwater}, {Freeman}, {Gibson}, {Green}, {Haynes}, {Juraszek},
  {Kesteven}, {Knezek}, {Mader}, {Marquarding}, {Meyer}, {Mould}, {Oosterloo},
  {O'Brien}, {Price}, {Sadler}, {Schr{\"o}der}, {Stewart}, {Stootman}, {Waugh},
  {Warren}, {Webster}, \& {Wright}}]{Koribalski2004AJ....128...16K}
{Koribalski}, B.~S., {Staveley-Smith}, L., {Kilborn}, V.~A., {et~al.} 2004,
  \aj, 128, 16

\bibitem[{{Kormendy} \& {Kennicutt}(2004)}]{Kormendy2004ARA&A..42..603K}
{Kormendy}, J. \& {Kennicutt}, Robert~C., J. 2004, \araa, 42, 603

\bibitem[{{Kumar} {et~al.}(2012){Kumar}, {Ghosh}, {Hutchings}, {Kamath},
  {Kathiravan}, {Mahesh}, {Murthy}, {Nagbhushana}, {Pati}, {Rao}, {Rao},
  {Sriram}, \& {Tandon}}]{Kumar2012SPIE.8443E..1NK}
{Kumar}, A., {Ghosh}, S.~K., {Hutchings}, J., {et~al.} 2012, in Society of
  Photo-Optical Instrumentation Engineers (SPIE) Conference Series, Vol. 8443,
  \procspie, 84431N

\bibitem[{{Lee} {et~al.}(2023){Lee}, {Sandstrom}, {Leroy}, {Thilker},
  {Schinnerer}, {Rosolowsky}, {Larson}, {Egorov}, {Williams}, {Schmidt},
  {Emsellem}, {Anand}, {Barnes}, {Belfiore}, {Be{\v{s}}li{\'c}}, {Bigiel},
  {Blanc}, {Bolatto}, {Boquien}, {den Brok}, {Cao}, {Chandar}, {Chastenet},
  {Chevance}, {Chiang}, {Congiu}, {Dale}, {Deger}, {Eibensteiner}, {Faesi},
  {Glover}, {Grasha}, {Groves}, {Hassani}, {Henny}, {Henshaw}, {Hoyer},
  {Hughes}, {Jeffreson}, {Jim{\'e}nez-Donaire}, {Kim}, {Kim}, {Klessen},
  {Koch}, {Kreckel}, {Kruijssen}, {Li}, {Liu}, {Lopez}, {Maschmann}, {Chen},
  {Meidt}, {Murphy}, {Neumann}, {Neumayer}, {Pan}, {Pessa}, {Pety},
  {Querejeta}, {Pinna}, {Rodr{\'\i}guez}, {Saito}, {S{\'a}nchez-Bl{\'a}zquez},
  {Santoro}, {Sardone}, {Smith}, {Sormani}, {Scheuermann}, {Stuber}, {Sutter},
  {Sun}, {Teng}, {Tre{\ss}}, {Usero}, {Watkins}, {Whitmore}, \&
  {Razza}}]{Lee2023ApJ...944L..17L}
{Lee}, J.~C., {Sandstrom}, K.~M., {Leroy}, A.~K., {et~al.} 2023, \apjl, 944,
  L17

\bibitem[{{Lelli} {et~al.}(2014){Lelli}, {Verheijen}, \&
  {Fraternali}}]{Lelli2014MNRAS.445.1694L}
{Lelli}, F., {Verheijen}, M., \& {Fraternali}, F. 2014, \mnras, 445, 1694

\bibitem[{{Leroy} {et~al.}(2008){Leroy}, {Walter}, {Brinks}, {Bigiel}, {de
  Blok}, {Madore}, \& {Thornley}}]{Leroy2008AJ....136.2782L}
{Leroy}, A.~K., {Walter}, F., {Brinks}, E., {et~al.} 2008, \aj, 136, 2782

\bibitem[{{Li} {et~al.}(2008){Li}, {Kauffmann}, {Heckman}, {White}, \&
  {Jing}}]{Li2008}
{Li}, C., {Kauffmann}, G., {Heckman}, T.~M., {White}, S. D.~M., \& {Jing},
  Y.~P. 2008, \mnras, 385, 1915

\bibitem[{{Lord} \& {Kenney}(1991)}]{Lord_BARCROWDING1991ApJ...381..130L}
{Lord}, S.~D. \& {Kenney}, J. D.~P. 1991, \apj, 381, 130

\bibitem[{López-Sánchez {et~al.}(2015)López-Sánchez, Westmeier, Esteban, \&
  Koribalski}]{Lopez10.1093/mnras/stv703}
López-Sánchez, A.~R., Westmeier, T., Esteban, C., \& Koribalski, B.~S. 2015,
  Monthly Notices of the Royal Astronomical Society, 450, 3381

\bibitem[{{Ma} {et~al.}(2017){Ma}, {de Grijs}, \& {Ho}}]{Ma2017ApJS..230...14M}
{Ma}, C., {de Grijs}, R., \& {Ho}, L.~C. 2017, \apjs, 230, 14

\bibitem[{{Maeda} {et~al.}(2023){Maeda}, {Egusa}, {Ohta}, {Fujimoto}, \&
  {Habe}}]{Maeda2023ApJ...943....7M}
{Maeda}, F., {Egusa}, F., {Ohta}, K., {Fujimoto}, Y., \& {Habe}, A. 2023, \apj,
  943, 7

\bibitem[{{Malin} \& {Hadley}(1997)}]{Malin1997PASA...14...52M}
{Malin}, D. \& {Hadley}, B. 1997, \pasa, 14, 52

\bibitem[{{Maoz} {et~al.}(2001){Maoz}, {Barth}, {Ho}, {Sternberg}, \&
  {Filippenko}}]{Maoz2001AJ....121.3048M}
{Maoz}, D., {Barth}, A.~J., {Ho}, L.~C., {Sternberg}, A., \& {Filippenko},
  A.~V. 2001, \aj, 121, 3048

\bibitem[{{Martel} {et~al.}(2013){Martel}, {Kawata}, \&
  {Ellison}}]{Martel2013MNRAS.431.2560M}
{Martel}, H., {Kawata}, D., \& {Ellison}, S.~L. 2013, \mnras, 431, 2560

\bibitem[{{Martin} {et~al.}(2005){Martin}, {Fanson}, {Schiminovich},
  {Morrissey}, {Friedman}, {Barlow}, {Conrow}, {Grange}, {Jelinsky},
  {Milliard}, {Siegmund}, {Bianchi}, {Byun}, {Donas}, {Forster}, {Heckman},
  {Lee}, {Madore}, {Malina}, {Neff}, {Rich}, {Small}, {Surber}, {Szalay},
  {Welsh}, \& {Wyder}}]{martingalex2005}
{Martin}, D.~C., {Fanson}, J., {Schiminovich}, D., {et~al.} 2005, \apjl, 619,
  L1

\bibitem[{{Mart{\'\i}nez-Delgado} {et~al.}(2010){Mart{\'\i}nez-Delgado},
  {Gabany}, {Crawford}, {Zibetti}, {Majewski}, {Rix}, {Fliri},
  {Carballo-Bello}, {Bardalez-Gagliuffi}, {Pe{\~n}arrubia}, {Chonis}, {Madore},
  {Trujillo}, {Schirmer}, \& {McDavid}}]{Delgado2010AJ....140..962M}
{Mart{\'\i}nez-Delgado}, D., {Gabany}, R.~J., {Crawford}, K., {et~al.} 2010,
  \aj, 140, 962

\bibitem[{{Mart{\'\i}nez-Delgado} {et~al.}(2009){Mart{\'\i}nez-Delgado},
  {Pohlen}, {Gabany}, {Majewski}, {Pe{\~n}arrubia}, \&
  {Palma}}]{Delgado2009ApJ...692..955M}
{Mart{\'\i}nez-Delgado}, D., {Pohlen}, M., {Gabany}, R.~J., {et~al.} 2009,
  \apj, 692, 955

\bibitem[{{Mathewson} \& {Ford}(1996)}]{Mathewson1996ApJS..107...97M}
{Mathewson}, D.~S. \& {Ford}, V.~L. 1996, \apjs, 107, 97

\bibitem[{{Meurer} {et~al.}(2006){Meurer}, {Hanish}, {Ferguson}, {Knezek},
  {Kilborn}, {Putman}, {Smith}, {Koribalski}, {Meyer}, {Oey}, {Ryan-Weber},
  {Zwaan}, {Heckman}, {Kennicutt}, {Lee}, {Webster}, {Bland-Hawthorn},
  {Dopita}, {Freeman}, {Doyle}, {Drinkwater}, {Staveley-Smith}, \&
  {Werk}}]{Meurer2006ApJS..165..307M}
{Meurer}, G.~R., {Hanish}, D.~J., {Ferguson}, H.~C., {et~al.} 2006, \apjs, 165,
  307

\bibitem[{{Offringa} {et~al.}(2014){Offringa}, {McKinley}, \&
  {Hurley-Walker}}]{Offringa2014MNRAS.444..606O}
{Offringa}, A.~R., {McKinley}, B., \& {Hurley-Walker}, N. 2014, \mnras, 444,
  606

\bibitem[{Offringa {et~al.}(2012)Offringa, van~de Gronde, \&
  Roerdink}]{offringa-2012-morph-rfi-algorithm}
Offringa, A.~R., van~de Gronde, J.~J., \& Roerdink, J. B. T.~M. 2012, A\&A, 539

\bibitem[{{Parkash} {et~al.}(2018){Parkash}, {Brown}, {Jarrett}, \&
  {Bonne}}]{Parkash2018ApJ...864...40P}
{Parkash}, V., {Brown}, M. J.~I., {Jarrett}, T.~H., \& {Bonne}, N.~J. 2018,
  \apj, 864, 40

\bibitem[{{Peters} \& {Kuzio de Naray}(2018)}]{Peters2018MNRAS.476.2938P}
{Peters}, W. \& {Kuzio de Naray}, R. 2018, \mnras, 476, 2938

\bibitem[{Postma \& Leahy(2017)}]{Postma_Leahy_2017}
Postma, J.~E. \& Leahy, D. 2017, Publications of the Astronomical Society of
  the Pacific, 129, 115002

\bibitem[{Pratt \& Gibbons(1981)}]{pratt1981kolmogorov}
Pratt, J.~W. \& Gibbons, J.~D. 1981, in Concepts of nonparametric theory
  (Springer), 318--344

\bibitem[{{Renaud} {et~al.}(2015){Renaud}, {Bournaud}, {Emsellem}, {Agertz},
  {Athanassoula}, {Combes}, {Elmegreen}, {Kraljic}, {Motte}, \&
  {Teyssier}}]{Renaud2015MNRAS.454.3299R}
{Renaud}, F., {Bournaud}, F., {Emsellem}, E., {et~al.} 2015, \mnras, 454, 3299

\bibitem[{{Renaud} {et~al.}(2013){Renaud}, {Bournaud}, {Emsellem}, {Elmegreen},
  {Teyssier}, {Alves}, {Chapon}, {Combes}, {Dekel}, {Gabor}, {Hennebelle}, \&
  {Kraljic}}]{Renaud2013MNRAS.436.1836R}
{Renaud}, F., {Bournaud}, F., {Emsellem}, E., {et~al.} 2013, \mnras, 436, 1836

\bibitem[{Robotham {et~al.}(2018)Robotham, Davies, Driver, Koushan, Taranu,
  Casura, \& Liske}]{Robotham_Davies_Driver_Koushan_Taranu_Casura_Liske_2018}
Robotham, A. S.~G., Davies, L. J.~M., Driver, S.~P., {et~al.} 2018, Monthly
  Notices of the Royal Astronomical Society, 476, 3137–3159

\bibitem[{{Saha} {et~al.}(2021){Saha}, {Dhiwar}, {Barway}, {Narayan}, \&
  {Tandon}}]{Saha2021JApA...42...59S}
{Saha}, K., {Dhiwar}, S., {Barway}, S., {Narayan}, C., \& {Tandon}, S. 2021,
  Journal of Astrophysics and Astronomy, 42, 59

\bibitem[{{Schlafly} \& {Finkbeiner}(2011)}]{Schlafly2011}
{Schlafly}, E.~F. \& {Finkbeiner}, D.~P. 2011, \apj, 737, 103

\bibitem[{{Schwarz}(1981)}]{Schwarz1981ApJ...247...77S}
{Schwarz}, M.~P. 1981, \apj, 247, 77

\bibitem[{{Schwarz}(1984)}]{Schwarz1984MNRAS.209...93S}
{Schwarz}, M.~P. 1984, \mnras, 209, 93

\bibitem[{{Serra} {et~al.}(2015){Serra}, {Westmeier}, {Giese}, {Jurek},
  {Fl{\"o}er}, {Popping}, {Winkel}, {van der Hulst}, {Meyer}, {Koribalski},
  {Staveley-Smith}, \& {Courtois}}]{Serra2015MNRAS.448.1922S}
{Serra}, P., {Westmeier}, T., {Giese}, N., {et~al.} 2015, \mnras, 448, 1922

\bibitem[{{Skibba} {et~al.}(2009){Skibba}, {Bamford}, {Nichol}, {Lintott},
  {Andreescu}, {Edmondson}, {Murray}, {Raddick}, {Schawinski}, {Slosar},
  {Szalay}, {Thomas}, \& {Vandenberg}}]{Skibba2009MNRAS.399..966S}
{Skibba}, R.~A., {Bamford}, S.~P., {Nichol}, R.~C., {et~al.} 2009, \mnras, 399,
  966

\bibitem[{{Smirnova} {et~al.}(2020){Smirnova}, {Wiebe}, {Moiseev}, \&
  {Jozsa}}]{Smirnova2020AstBu..75..234S}
{Smirnova}, K.~I., {Wiebe}, D.~S., {Moiseev}, A.~V., \& {Jozsa}, G.~I.~G. 2020,
  Astrophysical Bulletin, 75, 234

\bibitem[{{Struck}(1999)}]{Struck1999PhR...321....1S}
{Struck}, C. 1999, \physrep, 321, 1

\bibitem[{{Struck} {et~al.}(2011){Struck}, {Dobbs}, \&
  {Hwang}}]{Struck2011MNRAS.414.2498S}
{Struck}, C., {Dobbs}, C.~L., \& {Hwang}, J.-S. 2011, \mnras, 414, 2498

\bibitem[{{Tandon} {et~al.}(2020){Tandon}, {Postma}, {Joseph}, {Devaraj},
  {Subramaniam}, {Barve}, {George}, {Ghosh}, {Girish}, {Hutchings}, {Kamath},
  {Kathiravan}, {Kumar}, {Lancelot}, {Leahy}, {Mahesh}, {Mohan},
  {Nagabhushana}, {Pati}, {Rao}, {Sankarasubramanian}, {Sriram}, \&
  {Stalin}}]{Tandon2020AJ....159..158T}
{Tandon}, S.~N., {Postma}, J., {Joseph}, P., {et~al.} 2020, \aj, 159, 158

\bibitem[{{Tandon} {et~al.}(2017){Tandon}, {Subramaniam}, {Girish}, {Postma},
  {Sankarasubramanian}, {Sriram}, {Stalin}, {Mondal}, {Sahu}, {Joseph},
  {Hutchings}, {Ghosh}, {Barve}, {George}, {Kamath}, {Kathiravan}, {Kumar},
  {Lancelot}, {Leahy}, {Mahesh}, {Mohan}, {Nagabhushana}, {Pati}, {Kameswara
  Rao}, {Sreedhar}, \& {Sreekumar}}]{Tandon2017}
{Tandon}, S.~N., {Subramaniam}, A., {Girish}, V., {et~al.} 2017, \aj, 154, 128

\bibitem[{{Thilker} {et~al.}(2007){Thilker}, {Bianchi}, {Meurer}, {Gil de Paz},
  {Boissier}, {Madore}, {Boselli}, {Ferguson}, {Mu{\~n}oz-Mateos}, {Madsen},
  {Hameed}, {Overzier}, {Forster}, {Friedman}, {Martin}, {Morrissey}, {Neff},
  {Schiminovich}, {Seibert}, {Small}, {Wyder}, {Donas}, {Heckman}, {Lee},
  {Milliard}, {Rich}, {Szalay}, {Welsh}, \& {Yi}}]{Thilker2007ApJS..173..538T}
{Thilker}, D.~A., {Bianchi}, L., {Meurer}, G., {et~al.} 2007, \apjs, 173, 538

\bibitem[{Toomre \& Toomre(1972)}]{toomre1972galactic}
Toomre, A. \& Toomre, J. 1972, The Astrophysical Journal, 178, 623

\bibitem[{{Toomre} \& {Toomre}(1972)}]{Toomre1972}
{Toomre}, A. \& {Toomre}, J. 1972, \apj, 178, 623

\bibitem[{{Tully} {et~al.}(2016){Tully}, {Courtois}, \&
  {Sorce}}]{TullyDISTANCE2016AJ....152...50T}
{Tully}, R.~B., {Courtois}, H.~M., \& {Sorce}, J.~G. 2016, \aj, 152, 50

\bibitem[{{Ujjwal} {et~al.}(2022){Ujjwal}, {Kartha}, {Subramanian}, {George},
  {Thomas}, \& {Mathew}}]{Ujjwal2022MNRAS.tmp.2176U}
{Ujjwal}, K., {Kartha}, S.~S., {Subramanian}, S., {et~al.} 2022, \mnras, 516,
  2171

\bibitem[{{Urquhart} {et~al.}(2021){Urquhart}, {Figura}, {Cross}, {Wells},
  {Moore}, {Eden}, {Ragan}, {Pettitt}, {Duarte-Cabral}, {Colombo}, {Schuller},
  {Csengeri}, {Mattern}, {Beuther}, {Menten}, {Wyrowski}, {Anderson}, {Barnes},
  {Beltr{\'a}n}, {Billington}, {Bronfman}, {Giannetti}, {Kainulainen},
  {Kauffmann}, {Lee}, {Leurini}, {Medina}, {Montenegro-Montes}, {Riener},
  {Rigby}, {S{\'a}nchez-Monge}, {Schilke}, {Schisano}, {Traficante}, \&
  {Wienen}}]{Urquhart2021MNRAS.500.3050U}
{Urquhart}, J.~S., {Figura}, C., {Cross}, J.~R., {et~al.} 2021, \mnras, 500,
  3050

\bibitem[{{van der Marel} \& {Cioni}(2001)}]{Marel2001AJ....122.1807V}
{van der Marel}, R.~P. \& {Cioni}, M.-R.~L. 2001, \aj, 122, 1807

\bibitem[{{Zhang} {et~al.}(2020){Zhang}, {Smith}, {Oh}, {Paudel}, {Duc},
  {Boselli}, {C{\^o}t{\'e}}, {Ferrarese}, {Gao}, {Hunter}, {Puzia}, {Peng},
  {Rong}, {Shin}, \& {Zhao}}]{Zhang2020ApJ...900..152Z}
{Zhang}, H.-X., {Smith}, R., {Oh}, S.-H., {et~al.} 2020, \apj, 900, 152

\bibitem[{{Zhou} {et~al.}(2018){Zhou}, {Wu}, {Zhou}, \&
  {Ma}}]{Zhou2018PASP..130i4101Z}
{Zhou}, Z., {Wu}, H., {Zhou}, X., \& {Ma}, J. 2018, \pasp, 130, 094101

\end{thebibliography}

\end{document}